\documentclass[]{aa}
\usepackage[colorlinks=true,linkcolor=blue,citecolor=blue]{hyperref}
\usepackage[dvipsnames]{xcolor}
\usepackage{graphicx}
\graphicspath{{graphics}}
\usepackage{txfonts}
\usepackage{upgreek}
\usepackage[separate-uncertainty=true, list-units=repeat]{siunitx}
\usepackage{multirow}
\usepackage{bm}

\usepackage{soul}
\usepackage{float,lscape}
\usepackage{subcaption}
\usepackage{amsmath}
\usepackage{physics} 	
\usepackage{color} 		
\usepackage{verbatim} 	
\usepackage{threeparttable} 
\usepackage{natbib}
\bibpunct{(}{)}{;}{a}{}{,} 

\newcommand{\JWST}{\textit{JWST}}

\DeclareSIUnit{\Msun}{M_\odot}
\DeclareSIUnit{\year}{yr}
\DeclareSIUnit{\pc}{pc}
\DeclareSIUnit{\mag}{mag}
\DeclareSIUnit{\mas}{mas}
\DeclareSIUnit{\dex}{dex}
\DeclareSIUnit{\jansky}{Jy}
\DeclareSIUnit{\kpc}{kpc}

\newcommand{\vect}[1]{\boldsymbol{#1}}
\newcommand{\sigmaUV}{$\sigma_{\rm UV}$}
\newcommand{\MUV}{$M_{\rm UV}$}
\newcommand{\Mh}{$M_{\rm h}$}

\def\ha{H$\upalpha$}

\def\oiii{[\ion{O}{III}]}

\begin{document} 

    \title{Constraints on the early Universe star formation efficiency from galaxy clustering and halo modeling of H$\upalpha$ and [\ion{O}{III}] emitters}
   \subtitle{}  

    \titlerunning{Early Universe SFE from clustering}
   \author{
    Marko Shuntov\inst{\ref{DAWN},\ref{NBI},\ref{UniGe}}\fnmsep\thanks{\email{marko.shuntov@nbi.ku.dk}} 
    \and
    Pascal A.~Oesch\inst{\ref{UniGe}, \ref{DAWN},\ref{NBI}} 
    \and%
    Sune Toft\inst{\ref{DAWN},\ref{NBI}} 
    \and
    Romain A. Meyer \inst{\ref{UniGe}}
    \and
    Alba Covelo-Paz\inst{\ref{UniGe}} 
    \and%
    Louise Paquereau\inst{\ref{IAP}} 
    \and
    Rychard Bouwens\inst{\ref{Leiden}}
    \and
    Gabriel Brammer\inst{\ref{DAWN}, \ref{NBI}}
    \and
    Viola Gelli\inst{\ref{DAWN}, \ref{NBI}} 
    \and
    Emma Giovinazzo\inst{\ref{UniGe}} 
    \and
    Thomas Herard-Demanche\inst{\ref{Leiden}}
    \and
    Garth D. Illingworth\inst{\ref{SCruz}}
    \and%
    Charlotte Mason\inst{\ref{DAWN}, \ref{NBI}} 
    \and
    Rohan P. Naidu\inst{\ref{MIT},\ref{HubbleFellow}}
    \and
    Andrea Weibel\inst{\ref{UniGe}} 
    \and
    Mengyuan Xiao\inst{\ref{UniGe}}
    }

   \institute{
    Cosmic Dawn Center (DAWN), Denmark\label{DAWN}%
    \and%
    Niels Bohr Institute, University of Copenhagen, Jagtvej 128, 2200 Copenhagen, Denmark \label{NBI}%
    \and%
    Department of Astronomy, University of Geneva, Chemin Pegasi 51, 1290 Versoix, Switzerland \label{UniGe} 
    \and
    Institut d’Astrophysique de Paris, UMR 7095, CNRS, Sorbonne Université, 98 bis boulevard Arago, F-75014 Paris, France\label{IAP}%
    \and
    Leiden Observatory, Leiden University, PO Box 9513, 2300 RA Leiden, the Netherlands \label{Leiden}
    \and
    Department of Astronomy and Astrophysics, University of California, Santa Cruz, CA 95064, USA \label{SCruz}
    \and
    MIT Kavli Institute for Astrophysics and Space Research, 77 Massachusetts Ave., Cambridge, MA 02139, USA \label{MIT}
    \and
    NASA Hubble Fellow \label{HubbleFellow}
   }

   \date{Received ; accepted }

 
  \abstract
    {
    We develop a theoretical framework necessary to provide observational constraints on the early Universe galaxy-halo connection by combining measurements of the ultraviolet (UV) luminosity function (UVLF) and galaxy clustering via the 2-point correlation function (2PCF). 
    We implemented this framework in the FRESCO and CONGRESS \JWST\ NIRCam/grism  surveys by measuring the 2PCF of spectroscopically selected samples of \ha\ and \oiii \ emitters at $3.8<z<9$ in 124 arcmin$^2$ in GOODS-North and GOODS-South.
    By fitting the 2PCF and UVLF at $3.8<z<9$ we inferred that the \ha\ and \oiii\ samples at $\langle z \rangle \sim4.3, 5.4$ and $7.3$ reside in halos of masses of log$(M_{\rm h}/$M$_{\odot}) = 11.5$, $11.2$, $11.0$ respectively, while their galaxy bias increases with redshift with values of $b_{\rm g} = 4.0$, $5.0$, $7.6$. These halos, however, do not represent extreme overdense environments at these epochs.
    Our framework constrains the instantaneous star formation efficiency (SFE), defined as the ratio of the star formation rate over the baryonic accretion rate as a function of halo mass. We find that the SFE rises with halo mass, peaks at $\sim20\%$ at $M_{\rm h} \sim 3 \times 10^{11}\, \si{\Msun}$, and declines at higher halo masses. The SFE-$M_{\rm h}$ shows only a mild evolution with redshift with tentative indications that low mass halos decrease but the high mass halos increase in efficiency with redshift. The scatter in the $M_{\rm UV}-M_{\rm h}$ relation, quantified by $\sigma_{\rm UV}$, implies modest stochasticity in the UV luminosities of $\sim 0.7$ mag, relatively constant with redshift. Extrapolating our model to $z>9$ shows that a constant SFE-$M_{\rm h}$ fixed at $z=8$ cannot reproduce the observed UVLF and neither high maximum SFE nor high stochasticity alone can explain the high abundances of luminous galaxies seen by \JWST. Extending the analysis of the UVLF and 2PCF to $z>9$ measured from wider surveys will be crucial in breaking degeneracies between different physical mechanisms that can explain the high abundance of bright galaxies.
    }


   \maketitle
%

\section{Introduction}

One of the most pressing questions in extragalactic astronomy currently is explaining the surprisingly high abundances of bright and massive galaxies in the early Universe revealed by the James Webb Space Telescope (\JWST). Statistical functions such as the ultraviolet luminosity function (UVLF) and the stellar mass function (SMF) are key observational measurements that inform our models of galaxy formation and evolution in the early Universe. Measurements of both the UVLF and SMF from \JWST\ data (e.g., \citealt{Donnan2023}, \citealt{Harikane2023}, \citealt{Perez-Gonzale2023}, \citealt{Bouwens2023b}, \citealt{Finkelstein2023}, \citealt{Donnan2024}, \citealt{McLeod2024}, \citealt{Adams2024}, \citealt{Robertson2024} for the UVLF; \citealt{Weibel2024}, \citealt{Shuntov2024}, \citealt{Harvey2024}, \citealt{TWang2024} for the SMF) have revealed galaxy abundances in excess of theoretical models and simulations calibrated on the ``pre-\JWST\ Universe'' \citep[e.g.,][]{Boylan-Kolchin2023}.

To reconcile the theory with the observations, several physical mechanisms have been proposed. These can be broadly separated in two classes: 1) mechanisms that increase the star formation efficiency (SFE) in the early Universe, such as feedback-free starbursts \citep[FFB,][]{Torrey2017, Grudic2018, Dekel2023, Li2023, Renzini2023}; 2) mechanisms that do not need increased SFE and instead can produce abundant bright galaxies due to increased stochasticity in the UV luminosities \citep{Mason2023, Pallotini2023, Shen2023, Gelli2024}, via limited-to-no dust attenuation \citep{Ferrara2023} or a top-heavy initial mass function \citep{Cueto2023, Hutter2024} to name a few. The increased stochasticity manifests in the dispersion in the relation between galaxy UV magnitude (\MUV) and halo mass (\Mh), quantified by $\sigma_{\rm UV}$. Bursty star formation histories naturally produce this stochasticity and have been identified both observationally via the scatter in the star forming main sequence \citep[e.g.,][]{Ciesla2024, Clarke2024, Cole2025} and \ha-to-UV luminosity ratios \citep[e.g.,][]{Endsley2024}, as well as in simulations \citep{Pallotini2023, Sun2023, Muratov2015, Sparre2017}; although halo assembly histories and varying dust attenuation can also contribute to \sigmaUV\ \citep{Shen2023}.

Most of these models have been developed and calibrated to reproduce one-point statistical observables such as the UVLF and the UV luminosity density. However, the galaxy-halo connection is complex \citep{wechsler_connection_2018}, and there are several different mechanisms that can produce the same one-point observable, therefore making it intrinsically degenerate \citep{Munoz2023}. Crucial information lies in how galaxies are spatially distributed, not only in how many there are. The galaxy spatial distribution, or clustering, is described by two-point statistical functions, such as the two-point correlation function (2PCF). A successful galaxy formation model has to explain both the abundances and clustering of galaxies, therefore combining such complementary observables is key in breaking degeneracies between models \citep{Mirocha2020, Munoz2023, Gelli2024}.

On the observational side, using clustering as a probe alongside galaxy abundances has been done out to $z\sim7$ using Ly$\alpha$ emitters and Lyman-break galaxies (LBG) from Subary Suprime-Cam and Hyper Suprime-Cam (HSC) observations \citep{Ouchi2010, Ouchi2018, Harikane18, Harikane2022}.
\JWST\ has allowed measurements of the 2PCF out to the highest redshifts $z\sim10$ from photometrically selected samples \citep{Dalmasso24, Paquereau2025}. However, \JWST's NIRCam/grism capabilities enable us for the first time to compile complete spectroscopically selected samples out to $z\sim9$ \citep[e.g.][]{Oesch2023, Kashino2023} and measure their clustering \cite[as demonstrated by][for a limited sample of grism-selected \oiii-emitting galaxies at $5\lesssim z \lesssim7$ in a quasar field]{Eilers2024}.

On the theory side, the degeneracy issue has been explored by \cite{Mirocha2020, Munoz2023} who have shown that the two scenarios of high SFE versus high stochasticity that result in the same UVLF, have a very distinct impact on the galaxy bias. The latter is typically measured by comparing the observed 2PCF of galaxies to the theoretical one of dark matter halos, and describes how galaxies are preferentially distributed in the highest peaks of the underlying matter distribution. However, a comprehensive framework that models both the UVLF and 2PCF and can be used to derive empirical constrains of the theoretical models is still lacking.

In this paper, we develop a theoretical framework necessary to provide observational constraints on the early Universe galaxy-halo connection by combining measurements of the UVLF and galaxy clustering via the 2PCF. This framework is based on an adaptation of the halo occupation distribution (HOD) in combination with the conditional luminosity function and a parametric form of the SFE-\Mh\ relation. By fitting the observed UVLF and 2PCF in an independent binning scheme, it allows us to provide observational constrains on the \MUV-\Mh, and SFE-\Mh\ relations, stochasticity, host halo masses and galaxy bias. To implement this framework, we measured the 2PCF for the first time of spectroscopically selected samples of \ha\ and \oiii\ emitters at $3.8 < z <9$ from the \JWST\ NIRCam/grism surveys FRESCO \citep{Oesch2023} and CONGRESS \citep{Congress}. 

This paper is organized as follows. Section~\ref{sec:theoretical-framework} presents in detail the theoretical framework; Section~\ref{sec:data} presents briefly the NIRCam/grism data and methods used to compile the line emitting samples; Section~\ref{sec:measurements} describes the methods we used to measure the 2PCF and UVLF; Section~\ref{sec:results} presents the results of this work; In Section~\ref{sec:discussion} we discuss the physical implications of our results; and in Section~\ref{sec:conclussions} we summarize and conclude.

We adopt a standard $\Lambda$CDM cosmology with $H_0=70$\,km\,s$^{-1}$\,Mpc$^{-1}$ and $\Omega_{\rm m,0}=0.3$, where $\Omega_{\rm b,0}=0.04$, $\Omega_{\Lambda,0}=0.7$, and $\sigma_8 = 0.82$. All magnitudes are expressed in the AB system \citep{1983ApJ...266..713O}. Wherever relevant we assume a \cite{Salpeter1955} IMF and rescale literature datapoints to the same IMF.
The dependence of various quantities on the reduced Hubble parameter $h$ is retained implicitly. Dark matter halo masses, scale as $h^{-1}$. Absolute magnitude scale as $-5\,{\rm log}(h)$ in addition. When comparing to the literature, we rescale all the measurements to the cosmology adopted for this paper.

\section{Theoretical framework} \label{sec:theoretical-framework}

We develop a theoretical framework that is based on physically motivated principles and the halo model and from which we can derive fundamental statistical observables of galaxies such as the UVLF and the galaxy correlation function. At the heart of this formalism is a parametric model of the instantaneous star formation efficiency that relates the halo accretion rate to star formation, coupled with an HOD framework to predict galaxy clustering and abundances as a function of \MUV. 
This framework is analogous to the one developed to predict clustering and the galaxy stellar mass function by parametrizing the stellar-to-halo-mass relation \citep[e.g.,][]{Moster2010, Leauthaud2011, leauthaud_new_2012, Shuntov2022}. The novelty of this work is the extension to the UVLF, which also sets it apart from other works that use the 2PCF in combination to the total number density of \MUV-threshold selected galaxies \citep[e.g.,][]{Harikane18, Harikane2022}.

\subsection{The UV luminosity to halo mass relationship} \label{sec:uvhmr}

Galaxy star formation and mass growth is directly connected to the growth rate of dark matter halos, since the gas available for star formation is well mixed with the dark matter with a ratio of $f_{\rm b}\approx 0.16$ \citep[e.g.,][]{Moster2018, Tacchella2018}. This relation can be written as 
\begin{equation} \label{eq:sfr-mhdot}
    \dot{M}_{\star} = \epsilon \, f_{\rm b} \, \dot{M}_{\rm h}.
\end{equation}
The halo accretion rate, $\dot{M}_{\rm h}$, is well understood from $\Lambda$CDM, and we take the \cite{Dekel2013} form as a function of halo mass and redshift. The star formation rate $\dot{M}_{\star}$ is moderated by the instantaneous star formation efficiency (SFE) $\epsilon$. This function encapsulates all baryonic physics that is responsible for regulating star formation in halos and is essentially a summary of our galaxy formation model. Typically, it depends on both redshift and halo mass $\epsilon(z, M_{\rm h})$ and we adopt the double power law parametrization as a function of \Mh\ \citep{Moster2010, Moster2018}
\begin{equation} \label{eq:sfe-parametrization}
    \epsilon(M_{\rm h}) = 2\, \epsilon_0 \left[ \left(\dfrac{M_{\rm h}}{M_c}\right)^{-\beta} + \left(\dfrac{M_{\rm h}}{M_c}\right)^{\gamma} \right]^{-1}.
\end{equation}
This has four free parameters: $\epsilon_0$ -- the normalization, or the peak star formation efficiency; $M_{\rm c}$ -- a characteristic halo mass at which the efficiency peaks; $\beta$ and $\gamma$ -- the low- and high-mass slopes. In our implementation, we model these parameters as a function of redshift and fit for their redshift dependence \citep[following e.g.,][more details in \S\ref{sec:fitting-procedure}]{Moster2018, Munoz2023}.

We note here that Eq.~\ref{eq:sfr-mhdot} is distinct from the integrated star formation efficiency $\epsilon_{\star}$ that has been used frequently in recent high-$z$ literature \citep[e.g.,][]{BehrooziSilk2018, Boylan-Kolchin2023}. The latter quantifies the ratio between the stellar and halo mass assembled throughout the history of a halo, $\epsilon_{\star} = M_{\star} \, M_{\rm h}^{-1} \, f_{\rm b}^{-1}$, also known as the stellar-to-halo mass relationship \citep[SHMR,][]{wechsler_connection_2018}. The two are related by a time integral and do not necessarily have the same normalization and shape \citep{Behroozi13, Moster2018}, unless $\epsilon$ is Universal and time independent.

To arrive at the UV luminosity, we can use the fact that the SFR and UV luminosity are related with the luminosity-to-SFR conversion factor $\kappa_{\rm UV}$
\begin{equation} \label{eq:luminosity-from-sfr}
    L_{\rm UV} = \dfrac{\dot{M}_{\star}}{\kappa_{\rm UV}}.
\end{equation}
$\kappa_{\rm UV}$ is fully degenerate with $\epsilon_0$ \citep{Munoz2023}, but in a first and simplistic approach we will assume its fiducial value $\kappa_{\rm UV} = 1.15 \times 10^{-28} \, (M_{\odot}\, {\rm yr}^{-1})/({\rm erg}\, {\rm s}^{-1})$, for a Salpeter IMF \citep{MadauDickinson2014}. From the luminosity we can then derive the absolute magnitude $M_{\rm UV}$ in the AB system in the following way
\begin{equation} \label{eq:muv-from-luv}
    M_{\rm UV} = 51.63 - 2.5\,{\rm log} \left( {L_{\rm UV}/\mathrm{erg\,s^{-1}}}  \right).
\end{equation}
Finally, we add dust by using the $A_{\rm UV}-\beta$ relation from \cite{Meurer1999} and $\beta-M_{\rm UV}$ from \cite{Bouwens2014}, following the implementation in \cite{Mason2015, Sabti2022} and references therein. This assumes that $\beta$ follows a Gaussian distribution at each \MUV\, with dispersion $\sigma_{\beta} = 0.34$. The average dust extinction is then $\langle A_{\rm UV} \rangle = 4.43 + 0.79\, {\rm ln}(10)\, \sigma_{\beta} + 1.99 \, \langle \beta \rangle$.

The Eqs.~\ref{eq:sfr-mhdot}-\ref{eq:muv-from-luv} relate the UV luminosity $M_{\rm UV}$ to the halo mass $M_{\rm h}$, resulting in a $M_{\rm UV}$-$M_{\rm h}$ relationship, or UVHMR, which we denote $f_{\rm UVHMR}(M_{\rm h})$. \MUV\ has a monotonically increasing relation with \Mh\ that depends on redshift \citep[e.g.,][]{Mason2015}. The exact dependence is shaped by the change of both halo growth and the instantaneous SFE with time.

\subsection{The conditional UV luminosity function}

The \MUV-\Mh\ is not a one-to-one relation and there is a scatter in \MUV\ at a given \Mh. This can be modelled by a conditional luminosity function (CLF), that gives the average number of galaxies with \MUV$\pm \dd $\MUV$/2$ in a halo of mass \Mh\ \citep{Yang2003, Yang2009}. This can be modelled as a Gaussian distribution with a scatter $\sigma_{\rm UV}$
\begin{equation} \label{eq:CLF}
    P(M_{\rm UV}|M_{\rm h}) = \dfrac{1}{\sqrt{2\,\pi}\,\sigma_{\rm UV}} {\rm exp}\left[ -\left( \dfrac{M_{\rm UV} - f_{\rm UVHMR}(M_{\rm h})}{\sqrt{2} \, \sigma_{\rm UV}} \right)^2 \right],
\end{equation}
where $f_{\rm UVHMR}(M_{\rm h})$ gives the average \MUV\ as a function of halo mass $M_{\rm h}$. The scatter \sigmaUV\ is a result of intrinsic astrophysical processes, $\sigma_{\rm UV, intr}$, but also of  observational uncertainties in measuring \MUV, $\sigma_{\rm UV, meas}$ \citep{Leauthaud2011}, which can be written as $\sigma_{\rm UV}^2 = \sigma_{\rm UV, intr}^2 + \sigma_{\rm UV, meas}^2$. Additionally, \sigmaUV\ can be a function of halo mass \citep[e.g.][]{Sun2023, Gelli2024}, however in our implementation in this paper we simplify it and neglect this dependence on \Mh.

\subsection{The halo occupation distribution}

The CLF, crucially, provides a link to the HOD framework, which describes the statistical occupation of galaxies in dark matter halos and provides a formalism to model galaxy clustering via the two-point correlation function \citep[e.g.,][]{CooraySheth2002, berlind_halo_2002}. The HOD is a prescription of the mean number of galaxies residing in a halo of mass $M_{\rm h}$, which is given as
\begin{equation} \label{eq:hod-ntot}
    \left\langle N_{\rm tot}\left(M_{\rm h}\right) \right\rangle = \left\langle N_{\rm cent}\left(M_{\rm h}\right) \right\rangle \times \left[1 + \left\langle N_{\rm sat}\left(M_{\rm h}\right) \right\rangle\right].
\end{equation}
It is made of two components describing the mean occupation numbers of central and satellite galaxies. The occupation function for central galaxies brighter than some threshold luminosity $M_{\rm UV}^{\rm th}$, can be derived by integrating the CLF
\begin{equation} \label{eq:hod-ncent_formula}
    \left\langle N_{\rm cent}\left(M_{\rm h} | M_{\rm UV}^{\rm th} \right) \right\rangle = \displaystyle\int^{M_{\rm UV}^{\rm th}}_{\infty} {\rm d} M_{\rm UV} \, P(M_{\rm UV}|M_{\rm h})
\end{equation}
Carrying out this integration and accounting for the fact that \MUV\ is negative, we arrive at
\begin{equation} \label{eq:hod-ncent}
     \left\langle N_{\rm cent}\left(M_{\rm h} | M_{\rm UV}^{\rm th}\right) \right\rangle =
     \dfrac{1}{2} \left[ 1 - \text{erf} \left( 
    \dfrac{f_{\rm UVHMR}(M_{\rm h}) - M_{\rm UV}^{\rm th}}{\sqrt{2}\,\sigma_{{\rm UV}}} \right) \right].
\end{equation}
We can obtain this functional form by analytically integrating Eq.~\ref{eq:hod-ncent_formula}, thanks to the Gaussian form of $P(M_{\rm UV}|M_{\rm h})$, and because we assume a constant \sigmaUV. In principle, it is possible to allow a functional form of \sigmaUV\ with \Mh\ and carry out the integration of Eq.~\ref{eq:hod-ncent_formula} numerically to obtain $\langle N_{\rm cent} \rangle$. 

The occupation of halos by satellites can be modelled by a power-law for which we adopt the \cite{zheng_theoretical_2005} form
\begin{equation} \label{eq:hod-nsat}
    \left\langle N_{\rm sat}\left(M_h|M_{\rm UV}^{\rm th}\right) \right\rangle =
 \left( \dfrac{M_{\rm h} - M_{\rm cut}}{M_{\rm sat}} \right)^{\alpha_{\rm sat}},
\end{equation}
where $M_{\rm cut}$ and $M_{\rm sat}$ are the minimum and characteristic halo masses necessary to host satellite galaxies, and $\alpha_{\rm sat}$ is the power-law slope. This is different from other implementations of this formalism, where the satellite occupation is modeled by a combination of power- and exponential-law \citep{Leauthaud2011, coupon_galaxy-halo_2015, Shuntov2022}. We opt for this simpler description in order to reduce the number of free parameters.

One can also compute the average number of galaxies in a UV luminosity bin of $M_{\rm UV}^{\rm th1} < M_{\rm UV} < M_{\rm UV}^{\rm th2}$ by simply taking the difference of the occupation numbers for two \MUV\ thresholds
\begin{equation}
    \begin{split}
    & \left\langle N_{\rm cent/sat}\left(M_h|M_{\rm UV}^{\rm th1}, M_{\rm UV}^{\rm th2}\right) \right\rangle = \\
    & \left\langle N_{\rm cent/sat}\left(M_h|>M_{\rm UV}^{\rm th1}\right) \right\rangle - \left\langle N_{\rm cent/sat}\left(M_h|>M_{\rm UV}^{\rm th2}\right) \right\rangle.
    \end{split}
\end{equation}

\subsection{Model of the galaxy UV luminosity function} \label{sec:UVLF-model}
Having a description of the mean occupation numbers of galaxies as a function of their \MUV\ and of their host halo mass \Mh, we can derive an analytical form for the UV luminosity function (UVLF) by simply integrating over the halo mass function in the following way 
\begin{equation} 
        \Phi \left(M_{\rm UV}^{\rm th1}, M_{\rm UV}^{\rm th2}\right) =
        \displaystyle\int_{0}^{\infty} {\rm d} M_{\rm h} \left\langle N_{\rm tot}\left(M_{\rm h}|M_{\rm UV}^{\rm th1}, M_{\rm UV}^{\rm th2}\right)  \right\rangle \dfrac{{\rm d} n}{{\rm d} M_{\rm h}}.
\end{equation}
Here, $\Phi \left(M_{\rm UV}^{\rm th1}, M_{\rm UV}^{\rm th2}\right)$ indicates that the UVLF is computed for a \MUV\ bin defined by the two thresholds $M_{\rm UV}^{\rm th1}$ and $M_{\rm UV}^{\rm th2}$. For the integration limits, in practice we adopt the $[10^6-10^{15}]$ $M_{\odot}$ range. ${{\rm d} n}/{{\rm d} M_{\rm h}}$ is the halo mass function (HMF), for which we use the code \textsc{colossus} \citep{Diemer2018}, with the \cite{Watson2013} HMF and virial overdensity halo mass definition.

\subsection{Model of the 2-point correlation function} \label{sec:2pcf-model}

The model of the 2-point angular correlation function follows closely the usual prescriptions \citep[see for e.g.,][]{CooraySheth2002}. For completeness, we detail the principal equations in Appendix \ref{apdx:woftheta-model}. For the computation of the 2PCF, we rely on the \textsc{halomod}\footnote{\url{https://halomod.readthedocs.io/en/latest/}} code \citep{Murray2013, Murray2021}. \textsc{halomod} is a library of routines to calculate a range of cosmological observables. The main ingredients that enter the modeling along with the prescriptions and assumptions that we adopt are given in Table \ref{tab:assumptions}.

One of the ingredients in computing the 2PCF is the halo bias, which describes how halos trace the underlying dark matter distribution. Analytical forms of the halo bias are typically calibrated in the linear regime and on large scales using $N$-body simulations. However, not accounting for non-linear and scale-dependent effects fails to accurately describe the `quasi-linear' regime of the 1-halo to 2-halo transition ($r\sim10-500\, \si{\kpc}$) and can underpredict galaxy clustering by up to $30\%$ \citep{Reed09, Jose13, Mead15, Jose16, MeadVerde21}. As demonstrated in more detail in \cite{Harikane18, Paquereau2025}, accounting for non-linear and scale-dependent bias becomes important in accurately modeling the observed 2PCF at $z\gtrsim3$. To deal with this, we use the same implementation as in \cite{Paquereau2025}, based on the \cite{Jose16} non-linear scale-dependent corrections. Finally, we also apply the halo-exclusion correction that accounts for the fact that halos can not overlap and supresses the 2-halo power at scales smaller than the halo size \citep{Zheng2004, Tinker2005}.

\begin{table} [t] 
\centering
\caption{Adopted ingredients in the halo model}
\begin{tabular}{cc}
\hline
\hline
 Ingredient & Assumption  \\
 \hline
 HMF &  \cite{Watson2013}   \\ \\
 
 Halo bias $b_{\rm h}(M_{\rm h})$  &  \cite{tinker_large_2010}  \\ \\
 
 \shortstack{Non-linear scale-dependent\\halo bias correction} & \cite{Jose16} \\ \\
 
 \shortstack{Halo mass-\\concentration relation $c(M_h)$} &  \cite{Duffy08} \\ \\
 
 \shortstack{Halo and satellite\\over-density profiles} & \shortstack{NFW profile\\ \cite{NFW1997}} \\ \\
 
 Halo mass definition & virial \\
 \hline
 \hline
\end{tabular}
\label{tab:assumptions}
\end{table}

\subsection{Derived parameters} \label{sec:derived-params}
Using the mean halo occupation numbers and the halo mass function, we can derive various parameters related to the DM halos.
The number density of galaxies predicted by the HOD model is computed as
\begin{equation} \label{eq:ngal-hod}
    n_{\rm gal}(z) = \displaystyle\int \dd M_{\rm h} \, \dfrac{{\rm d} n}{{\rm d} M_{\rm h}}\, \langle N_{\rm tot} (M_{\rm h}) \rangle.
\end{equation} 
The average mass of the DM halo of galaxies at a given redshift is
\begin{equation} \label{eq:Mhalo}
    \langle M_{\rm h} \rangle(z) = \dfrac{1}{n_{\rm gal}(z)}\displaystyle\int \dd M_{\rm h} \, M_{\rm h} \, \dfrac{{\rm d} n}{{\rm d} M_{\rm h}}\, \langle N_{\rm tot} (M_{\rm h}) \rangle,
\end{equation}
The mean galaxy bias at a given redshift is
\begin{equation} \label{eq:bias}
    b_{\rm g} (z) = \dfrac{1}{ n_{\rm gal}(z)}\displaystyle\int \dd M_{\rm h} \, b_{\rm h}(M_{\rm h}, z) \, \dfrac{{\rm d} n}{{\rm d} M_{\rm h}}\, \langle N_{\rm tot} (M_{\rm h}) \rangle,
\end{equation}
where  $b_{\rm h}(M_{\rm h}, z)$ is the \cite{tinker_large_2010} linear halo bias.

\subsection{UVLF, 2PCF and galaxy bias dependence on model parameters}

In Appendix~\ref{apdx:model-param-depend}, Fig.~\ref{fig:model-param-dep} we show how the UVLF, 2PCF and galaxy bias change with different model parameter values. This demonstrates which measurements are most constraining for the model parameters. For this we adopt a set of fiducial parameter values $\sigma_{\rm UV} = 0.76, \, \epsilon_{0}=0.22, \, {\rm log} M_{\rm c}=11.56, \, \beta=0.69, \, \gamma=0.74, \, {\rm log} M_{\rm cut}=8.99, \, {\rm log} M_{\rm sat}=12.69, \alpha_{\rm sat}=0.76$, and we vary each parameter at a time, shown in the different rows in Fig.~\ref{fig:model-param-dep}.

The UVLF is most sensitive to the parameters that govern instantaneous SFE $\epsilon(M_{\rm h})$ and \sigmaUV. The scatter, \sigmaUV, mostly affects the bright end, such that a higher scatter flattens the bright-end slope, resulting in higher abundances of bright ($M_{\rm UV}<-21$) galaxies. Correspondingly, this reduces the clustering amplitude and galaxy bias. Increasing the maximum SFE, $\epsilon_0$ effectively shifts the UVLF towards the brighter end while decreasing the 2PCF amplitude and bias. This happens because higher efficiencies mean that lower mass halos, that are less clustered, can host brighter galaxies. The peak halo mass $M_{\rm c}$ governs the location of the `knee' of the UVLF with higher values moving the knee towards the brighter end, while increasing the 2PCF amplitude. The slopes $\beta$ and $\gamma$ affect the faint and bright end respectively, where increasing $\beta$, increases the 2PCF, while $\gamma$ has very little impact on the 2PCF. The parameters governing the satellite occupation ($M_{\rm sat}, \, M_{\rm cut}, \, \alpha_{\rm sat}$) have negligible impact on the UVLF and bias, but affect mostly the smaller scales of the 2PCF (1-halo term).

\subsection{Advantages and caveats}

The theoretical framework we developed for this work is highly flexible in modeling galaxy observables under a common parametrization of the instantaneous SFE, the \MUV$-$\Mh\ scatter, and the satellite occupation numbers. These model parameters can in turn be parametrized with redshift. This is a powerful approach that allows modeling the UVLF at any redshift bin under the same model parameters and consistently incorporate different $z$-bins under the same likelihood. Furthermore, this allows an independent binning scheme for the 2PCF and UVLF measurements. In fact, this is how we implement this framework in this paper (cf. \S\ref{sec:measurements-uvlf}).

Furthermore, this framework is highly modular -- the same parametrization can be used to model additional statistical observables such as the stellar mass function (SMF). This observable is complementary to the UVLF and subject to different systematics, and therefore can be a powerful addition to help break degeneracies and better constrain the model. We will explore this in a future work.

As detailed in \S\ref{sec:uvhmr}, the modeling includes a number of assumptions that can introduce systematic biases. For example, assumptions of the UV luminosity-to-SFR conversion factor $\kappa_{\rm UV}$ and the dust attenuation relations can be degenerate with the model parameters (we discuss this further in \S\ref{sec:discussion}). Furthermore, there is also a degeneracy with the duty cycle that gives the fraction of galaxies that are UV-bright at a given moment \citep[e.g.,][]{Mirocha2020, Munoz2023}. The duty cycle can be incorporated in the model by a multiplying the total occupation number $N_{\rm tot} \, f_{\rm duty}$ \citep[e.g.,][]{Harikane2016} and would result in scaling the 2PCF and UVLF differently \citep[][provide details on how exactly it affects the UVLF and galaxy bias]{Mirocha2020, Munoz2023}. For example, lowering $f_{\rm duty}$ would require increasing $\epsilon_0$ to match the UVLF, but would decrease the 2PCF amplitude.
We assume a duty cycle of unity, consistent with indications from previous work \citep{Harikane18, Harikane2022, Munoz2023}. Future work including larger datasets would allow us to model the duty cycle that can even be a function of halo mass \citep[e.g.,][]{Ouchi2001, Mirocha2020}.

\section{Data} \label{sec:data}

\begin{figure*}[t!]
\includegraphics[width=1\textwidth]{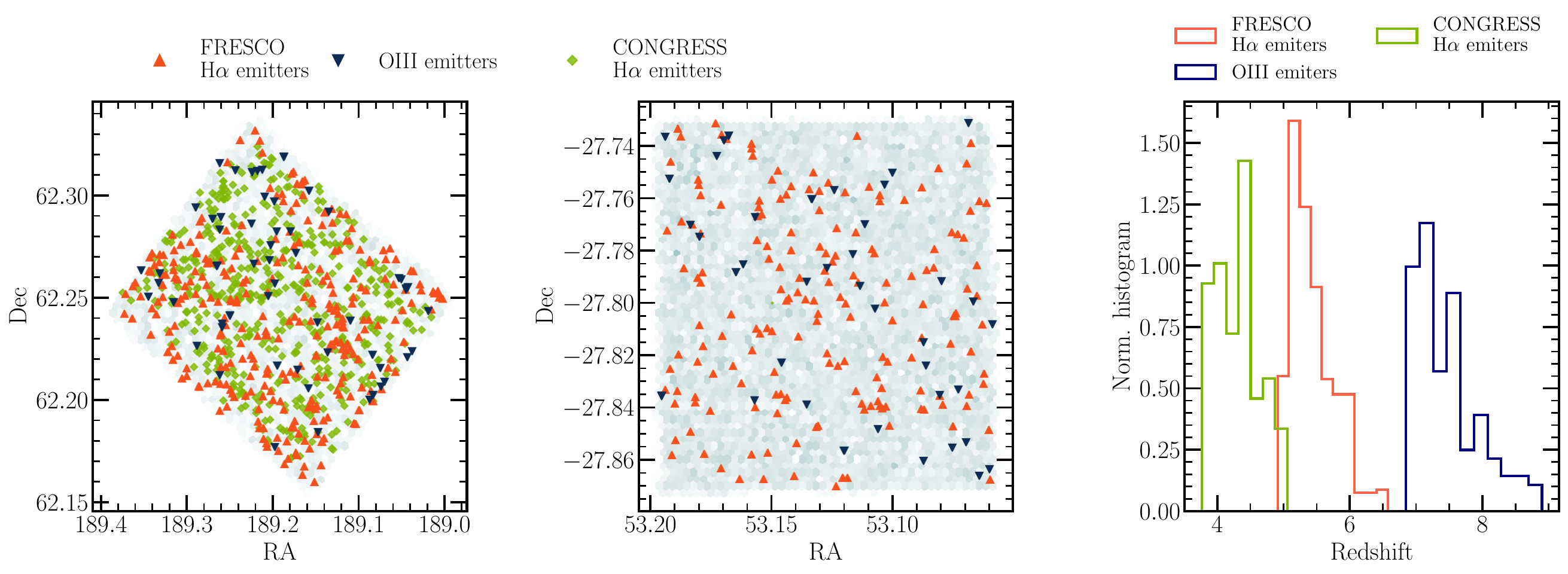}  
\caption{Spatial distribution of the three line-emitting samples (\ha\ and \oiii\ from FRESCO, and \ha\ from CONGRESS), in the GOODS-N and GOODS-S footprints (left and middle panels) which we consider in this analysis. The right panel shows their redshift distribution. The density histogram gray background in the left and middle panels shows the footprint of the random catalogs.}
\label{fig:source-distribution}
\end{figure*}

This work is anchored on the most comprehensive and complete \JWST NIRCam/grism spectroscopic surveys -- FRESCO \citep{Oesch2023} in the F444W filter and CONGRESS (Sun, Egami et al. in prep, \citealt{Congress}) in the F356W. These two surveys cover the $3.1 - 5.0\, \si{\micro \meter}$ range over $62 \, {\rm arcmin}^2$ in GOODS-North. In addition, FRESCO covers another $62 \, {\rm arcmin}^2$ in GOODS-South at $3.8 - 5.0\, \si{\micro \meter}$. Combined, these enable the compilation of complete samples of \ha\ emitters at $3.7<z<6.7$ ($4.9<z<6.7$) in GOODS-N (GOODS-S), and \oiii\ emitters at $6.8<z<9$.

We use the samples compiled by the works of \citet[][for \ha\ emitters]{Covelo-Paz2024} and \citet[][for \oiii\ emitters]{Meyer2024}; which we describe here briefly and refer to these references for more details. Photometric catalogs in the GOODS fields are extracted using all publicly available HST \citep[e.g.][in 10 bands]{Giavalisco2004,  Grogin2011, Koekemoer2011} and JWST \citep[e.g.][in 11 bands (19 for GOODS-S]{Eisenstein2023a, Eisenstein2023b, Williams2023}. Detection is carried out on an inverse-variance stack of the F210M and F444W bands from FRESCO to ensure homogenous depth.  Source extraction is done with \textsc{SExtractor} \citep{Bertin1996} on images that were PSF-matched to the F444W band. Aperture corrections are applied to scale the fluxes to those measured in Kron apertures on a PSF-matched version of the detection image, and then to total by dividing by the encircled energy of the Kron aperture on the F444W PSF. The detection completeness is estimated using the \textsc{GLACiAR2} software \citep{Leethochawalit2022} by measuring the fraction of injected-recovered sources as a function of F444W magnitude. The completeness is constant around 90\% out to 27 AB mag and declines to zero out to 30.5 AB mag. The NIRCam WFSS data is reduced and processed using the \textsc{grizli} software using the standard CRDS grism dispersion files from pmap 1123 \citep{Oesch2023, Meyer2024}. In the following, we briefly describe the selection of \ha\ and \oiii\ emitters.

\subsection{\ha\ sample} 
The \ha\ sample is constructed by first selecting all sources with \ha\ S/N$>3$ from the \textsc{grizli} catalog. Multi-component sources are identified upon visual inspection and components within $<0.6$ arcsec separation and identical grism redshifts are merged into one galaxy and assigned the same \ha\ line. Quality flags are assigned via visual inspection by four team members independently. We use quality flag $q\geq1.75$, which means that the source has a clear \ha\ line and a matching morphology between the direct image and the 2D-spectrum and the line map, or a lower S/N but a clear morphology map. Additionally, a high quality flag is assigned for FRESCO \ha\ emitter at $z>5.25$ that show a \oiii\ emission from the CONGRESS data \citep{Covelo-Paz2024}.

\subsection{\oiii\ sample} 
The \oiii\ sample is constructed by requiring S/N$\geq3$ in F444W FRESCO band and no detection in bands bluewards of Lyman-$\alpha$ to remove unambiguous low-$z$ contaminants. Then, for all these sources, the 1D spectra are filtered with a Gaussian with FWHM$=50,\, 100,\,200\, \si{\km \per \s}$ and candidates are retained if they satisfy the following conditions: 1) Two lines match the \oiii$\lambda\lambda5008,4960 \, \si{\AA}$ separation at $6.8 < z < 9.0$, with a tolerance of $\Delta\nu < 100\, \si{\km \per \s}$ on the doublet separation, 2) S/N$>4$ for the strongest line of the doublet, 3) The observed doublet ratio $1< \oiii5008/\oiii4960 < 10$. Visual inspection is carried out by multiple team members independently to remove contaminants and assign quality flags ($q$), similarly to the \ha\ catalog. We use objects with $q\geq1.5$, which is slightly less conservative than for the \ha\ sample; we opt for this lower value to maximize the number of objects for this intrinsically smaller sample of \oiii\ emitters \citep{Meyer2024}.

\subsection{Selection function}
The selection described above yields emission line-selected samples in three redshift bins -- two \ha\ emitter samples, the first at $3.7<z<5.1$ from CONGRESS, the second at $4.9<z<6.7$ from FRESCO, and a third sample of \oiii\ emitters at $6.8<z<9$. We apply a cut in the UV absolute magnitude to select only sources with $M_{\rm UV} < -19.1$. This threshold is chosen as the brightest magnitude that selects enough sources to allow measurement of the 2PCF $w(\theta)$. 
Practically, this requires that there is at least one pair of objects at the angular separations $\theta$ at which $w(\theta)$ is evaluated using the \cite{landy_bias_1993} estimator (\S\ref{sec:measurements-2pcf}, Eq.~\ref{eq:ls-estimator}). 

This selection function results in 398 galaxies for the $3.7<z<5.1$ \ha\ CONGRESS sample, 387 galaxies for the $4.9<z<6.7$ for the \ha\ FRESCO sample, and 108 galaxies for the $6.8<z<9$ \oiii\ sample. Fig.~\ref{fig:source-distribution} shows the spatial and redshift distributions of the three samples in the GOODS-N and GOODS-S fields.

\section{Measurements} \label{sec:measurements}

\subsection{Galaxy clustering via two-point angular correlation function} \label{sec:measurements-2pcf}

We measure galaxy clustering via the angular 2PCF, for the three emission line-selected samples, brighter than $M_{\rm UV} = -19.1$ mag, in the three redshift bins, using the \cite{landy_bias_1993} estimator\footnote{The correlation functions are computed using the \textsc{treecorr} code \citep{Jarvis2004}}. We measure the 2PCF using Eq.~\ref{eq:ls-estimator} for GOODS-North and GOODS-South separately and combine them together using a number density weighting scheme following \cite{durkalec_evolution_2015}
\begin{equation} \label{eq:ls-estimator}
    w(\theta) = \dfrac{\displaystyle\sum_{i=1}^{n_{\rm field}} w_{i}\, (DD_i - 2\,DR_i + RR_i) }{\displaystyle\sum_{i=1}^{n_{\rm field}} w_{i}\, RR_i},
\end{equation}
where $DD_i$, $RR_i$, $DR_i$ are the number of data-data, random-random, and data-random pairs in a given angular separation bin $[\theta, \theta+\delta \theta]$, normalized by the total number of galaxies and random objects for each field $i=1,2$. $w_i=(N_{i}/V_i)^2$ is the weight defined as the total number of galaxies devided by the volume of the corresponding field.

We construct random catalogs for the two fields and the two samples by Monte-Carlo sampling of the LF (\ha\ and \oiii\ accordingly) to assign a line flux to each random source. To account for the fact that the spatial selection function can be wavelength dependent we use the root-mean-square (RMS) cube given in \cite{Meyer2024} along with the completeness functions for the \oiii\ and \ha\ samples \citep{Meyer2024, Covelo-Paz2024}. We compute the $S/N$ for each source as a function of position and wavelength and retain the random draw with a probability equal to the completeness as a function of $S/N$. This ensures that the random catalog accounts for any spatial selection effects in the data.

Given the fact that this is a sample with highly precise redshifts from spectroscopy, it is in principle possible to compute the projected correlation as a function of comoving perpendicular separation $w_p(r_p)$. However, our samples are too small to carry out the measurement of $w_p(r_p)$ that is not too noisy for model fitting, and therefore we use the projected \emph{angular} correlation function $w(\theta)$ for our analysis. This is also consistent with the predictions by \cite{Endsley2020} using mock galaxy samples based on the \textsc{UniverseMachine} model.

Uncertainties in the 2PCF measurements come from Poisson pair-counting statistics and from cosmic variance \citep{Norberg2009}. The latter can have important impact on the measurements, especially for small field-of-view surveys like FRESCO. One way to estimate the cosmic-variance uncertainties is via jackknife, bootstrapping or Monte Carlo methods \citep{Norberg2009, Norberg2011}. However, the small number of sources and footprint render jackknife and bootstrapping methods unsuitable, while Monte Carlo methods would require cosmological simulations and synthetic galaxy catalogs, which is expensive and out of the scope of this work. Furthermore, using such simulations, \cite{Leauthaud2011} showed that cosmic variance has an impact on the covariance matrix on large scales (i.e. in the two-halo regime) -- it increases the correlation in the data at large scales, which is poorly constrained with our measurements due to the relatively small field.

We adopt a simplistic approach by considering only Poisson uncertainties, boosted by 30$\%$ in quadrature. Because of the small sample, it is also difficult to derive well-defined covariance matrices that are not too noisy for likelihood computation. For this reason, we use only the diagonal elements, and therefore not consider the covariance between different $\theta$ bins \citep[commonly adopted in the literature in case of small samples, e.g.][]{Zehavi2011}.

Due to the small volume probed in the GOODS-North and GOODS-South fields, the integral constraint (IC) suppresses/cuts $w(\theta)$ at scales near and beyond the survey size. We take this into account by adjusting the model with a correction factor that can be estimated from the double integration of the true correlation function over the survey area
\begin{equation}
    w_{\rm IC} = \dfrac{1}{\Omega^2} \displaystyle\int w_{\rm true} \, \dd \Omega_1 \dd \Omega_2.
\end{equation}
This integration can be carried out using the random-random pairs from the random catalog following \cite{roche_angular_1999}
\begin{equation}
    w_{\rm IC} = \dfrac{\sum w_{\rm true}(\theta) \, RR(\theta)}{\sum RR(\theta)},
\end{equation}
where $w_{\rm true} (\theta)$ comes from the HOD-predicted model. Finally, the model that we fit against the data is simply $w(\theta) = w_{\rm true} (\theta) - w_{\rm IC}$.

\subsection{UV Luminosity functions} \label{sec:measurements-uvlf}

Since our \ha\, and \oiii\, samples are identical to those used in \cite{Covelo-Paz2024} and \cite{Meyer2024}, we take the UVLF measurements presented in the respective papers. The completeness-corrected UVLFs are measured for the \ha\ CONGRESS at $z\sim4.3$, \ha\ FRESCO in two bins at $z\sim5.3$ and $z\sim5.6$, and the \oiii\ sample in two bins at $z \sim 7.1$ and $z\sim7.2$. The advantage of these UVLFs is that they are measured from secure spectroscopic samples, therefore alleviating systematic biases that can arise from typical photometrically selected samples. These are shown as yellow/red symbols in Fig.~\ref{fig:2pcf-uvlf-fits}.

However, these samples are limited at both the bright and faint end due the survey volume and sensitivity limits. In order to gain significant constraining power from the UVLFs, we also adopt the measurements from \cite{Bouwens2021} that use photometrically selected LBG over 1136 arcmin$^2$. We use the UVLFs in five bins at $z\sim4,5,6,7,8$ shown in gray symbols in Fig.~\ref{fig:2pcf-uvlf-fits}. As described in \S\ref{sec:UVLF-model}, our highly flexible modeling allows us to simultaneously model all of these ten UVLFs consistently and include them under the same likelihood. Finally, we boost the UVLF errorbars to account for about $25\%$ of cosmic variance uncertainty \citep[e.g.][]{Trapp2022, Sabti2022}.

\subsection{Fitting procedure} \label{sec:fitting-procedure}

We fit the measurements in all redshift bins simultaneously by assuming a parametric redshift dependence of the model parameters. In this way, we model a smooth redshift evolution of $\epsilon(M_{\rm h},z)$. We adopt a linear dependence with redshift for all parameters following \cite{Munoz2023}
\begin{equation} \label{eq:param-z-depend}
    X = \dfrac{\dd X}{\dd z}\,z + C(X),
\end{equation}
where $X = \left[\epsilon_{0}, \, {\rm log} M_{\rm c}, \, \beta, \, \gamma, \, \sigma_{\rm UV}, \, {\rm log} M_{\rm cut}, \, {\rm log} M_{\rm sat}, \alpha_{\rm sat} \right]$.

We fit the models of the $w(\theta)$ and the UVLFs to our measurements using a Markov Chain Monte Carlo (MCMC) approach, minimising $\chi^2$:
\begin{equation} \label{eq:likelihood}
\begin{split}
        \chi^2 = &   \displaystyle\sum_{i}^{N_{w}} (\vect{w}_i - \Tilde{\vect{w}}_i)^T C^{-1} (\vect{w}_i - \Tilde{\vect{w}}_i) \, + \\
        & \displaystyle\sum_{i}^{N_{\Phi}} \left( \dfrac{\Phi(M_{{\rm UV},i}) - \Tilde{\Phi}(M_{{\rm UV},i})}{\sigma_{\Phi}}\right)^2,
\end{split}
\end{equation}
where $\vect{w}$ are the measurement vectors containing $w$ at $\theta$, and $\Tilde{w}$ and $\Tilde{\Phi}$ are the models for a given set of parameter values. The first term of Eq. \ref{eq:likelihood} corresponds to the clustering likelihood and the second term to UVLF likelihood. The sums run over the different measurements in the three redshift bins for the clustering and 10 in total UVLFs at different redshifts (five for the line-emitter samples and five for the LBG).

We carry out the fitting using the \textsc{emcee} code \citep{foreman-mackey_emcee_2013}, that implements an affine-invariant ensemble sampler. We use $50$ walkers for our $5$ parameters and rely on the auto-correlation time $\tau$ to assess the convergence of the chain. To consider the chains converged, we require that the auto-correlation time is at least $50$ times the length of the chain and that the change in $\tau$ is less than $1\%$. We discard the first $2\times\text{max}(\tau)$ points of the chain as the burn-in phase and thin the resulting chain by $0.5\times\text{min}(\tau)$. We impose flat priors on all parameters; for the mass parameters, the flat priors are on the log quantities. 

For best-fit parameter values, we take the medians of the resulting posterior distribution, with the 16-th and 84-th percentiles giving the lower and upper uncertainty estimates. We compute the confidence intervals on all HOD model-derived quantities by computing it from 500 randomly drawn samples from the posterior, and taking the $1\,\sigma$ percentiles.
The posterior median and $1\,\sigma$ uncertainties of the main model parameters $X$ are given in Appendix~\ref{apdx:best-fit-params}, Table~\ref{tab:parameters}, for the three redshift bins of the \ha\ and \oiii\ samples. We also present the values for the parameters describing the linear with redshift dependence (Eq.~\ref{eq:param-z-depend}).

\section{Results} \label{sec:results}

\begin{figure*}[t!]
\includegraphics[width=1\textwidth]{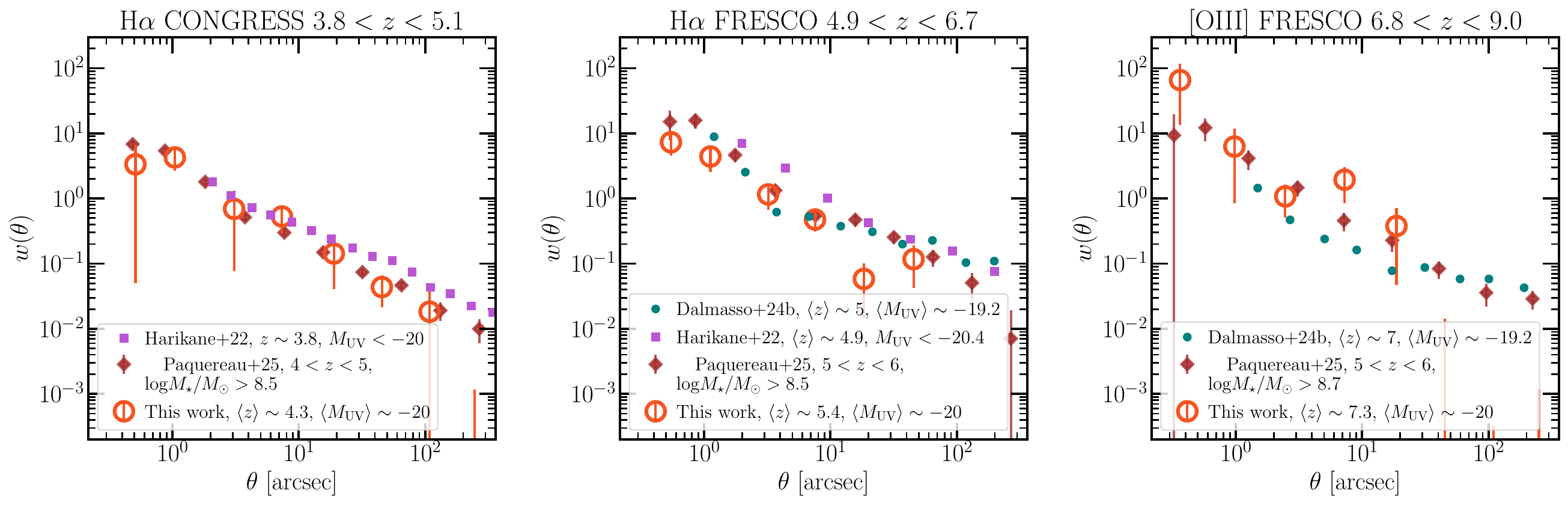}  
\caption{The angular 2PCF measured from the \ha\ CONGRESS and FRESCO samples at $z\sim4.3$ and $z\sim5.4$ and \oiii\ sample at $z\sim7.3$ selected at $M_{\rm UV} < -19.1$ mag. Comparison with literature measurements includes \cite{Harikane2022}, \cite{Dalmasso24}, and \cite{Paquereau2025} for samples selected at similar redshifts and \MUV\  as our work.}
\label{fig:2pcf-meas}
\end{figure*}

\subsection{Clustering of \ha\ and \oiii\ emitters} \label{sec:clustering-results}

We measure galaxy clustering via the angular 2PCF for spectroscopically selected galaxies out to $z\sim4.3$ and $z\sim5.4$ for \ha\ emitters, and $z\sim7.3$ for \oiii\ emitters brighter than $M_{\rm UV} = -19.1$ mag. These are shown in Fig.~\ref{fig:2pcf-meas}. The high resolution in NIRCam allows us to probe very small scales deep into the 1-halo regime, down to about 10 kpc at $z\sim6$; however, the limited survey area prevents us from probing scales larger than $\sim 1$ Mpc at $z\sim6$.

The amplitude of the 2PCF for the \ha\ emitters increases mildly from $z\sim4.3$ to $z\sim5.4$, and continues to increase for the \oiii\ emitters at $z\sim 7.3$. The shape of the 2PCF for the three samples resembles a power law, in contrast to the characteristic two-component behavior of the 1- and 2-halo terms. This is not surprising, given the fact that this is the regime where the nonlinear scale-dependent bias can amplify the power at intermediate $\sim100-500$ kpc scales \citep{Jose17}. Furthermore, relatively low-number statistics prevent us from accurately revealing the shape of the 2PCF for these samples. Finally, cosmic variance systematics (e.g., overdensities in the field) can have an influence on both the shape and amplitude of the 2PCF. Indeed, recent spectroscopic redshift searches for overdensities in the GOODS fields have revealed several significant overdensities at $z\sim6-7$ \citep{Helton2023, Meyer2024, Covelo-Paz2024, Herard-Demanche2025}. 

In Fig.~\ref{fig:2pcf-meas}, we also show a comparison with recent measurements from the literature. 
\cite{Dalmasso24} measure clustering of LBG in the JADES survey, and as such their sample overlaps with ours. We compare with their measurements at $z\sim5$ and $z\sim7$, but for fainter samples than ours with mean UV magnitudes of $\langle M_{\rm UV} \rangle \sim -19.2$, compared to ours $\langle M_{\rm UV} \rangle \sim -20$. In general, there is good agreement, with differences that are likely due to the redshift and \MUV\ selection; for example, at $z\sim7$ our 2PCF has expectedly larger amplitude because of brighter sample. 
\cite{Harikane2022} measure the 2PCF of photometrically selected LBG in the HSC Subaru Strategic Program (SSP) survey over $\sim300 \, \si{\deg}^2$ at $4<z<7$. There is relatively good agreement with slightly higher amplitude of their brighter, $M_{\rm UV} \lesssim -20$ sample, compared to our $M_{\rm UV} < -19.1$ mag sample.
Finally, we find excellent agreement with \cite{Paquereau2025} measurements in the $\sim0.5\, {\rm deg}^2$ COSMOS-Web survey for stellar mass selected samples of log$(M_{\star}/\si{\Msun})>8.5$. This is comparable to our \MUV\ threshold of $-19.1$ mag assuming the \cite{Song2016} mass-to-light ratio. Slight differences likely arise from the redshift selection.

This consistency of the 2PCF with the literature is indeed reassuring and showcases that photo-$z$ selected samples can yield robust and unbiased 2PCF, with an advantage of compiling significantly larger samples than what is possible with spec-$z$. This means that future implementation of our modeling framework will be possible using 2PCF measured from photo-$z$ selection, therefore reducing errorbars and extending the analysis to $z>9$.


\subsection{UVHMR $\times$ HOD model fitting} \label{sec:model-fitting-results}

\begin{figure*}[t!]
\includegraphics[width=1\textwidth]{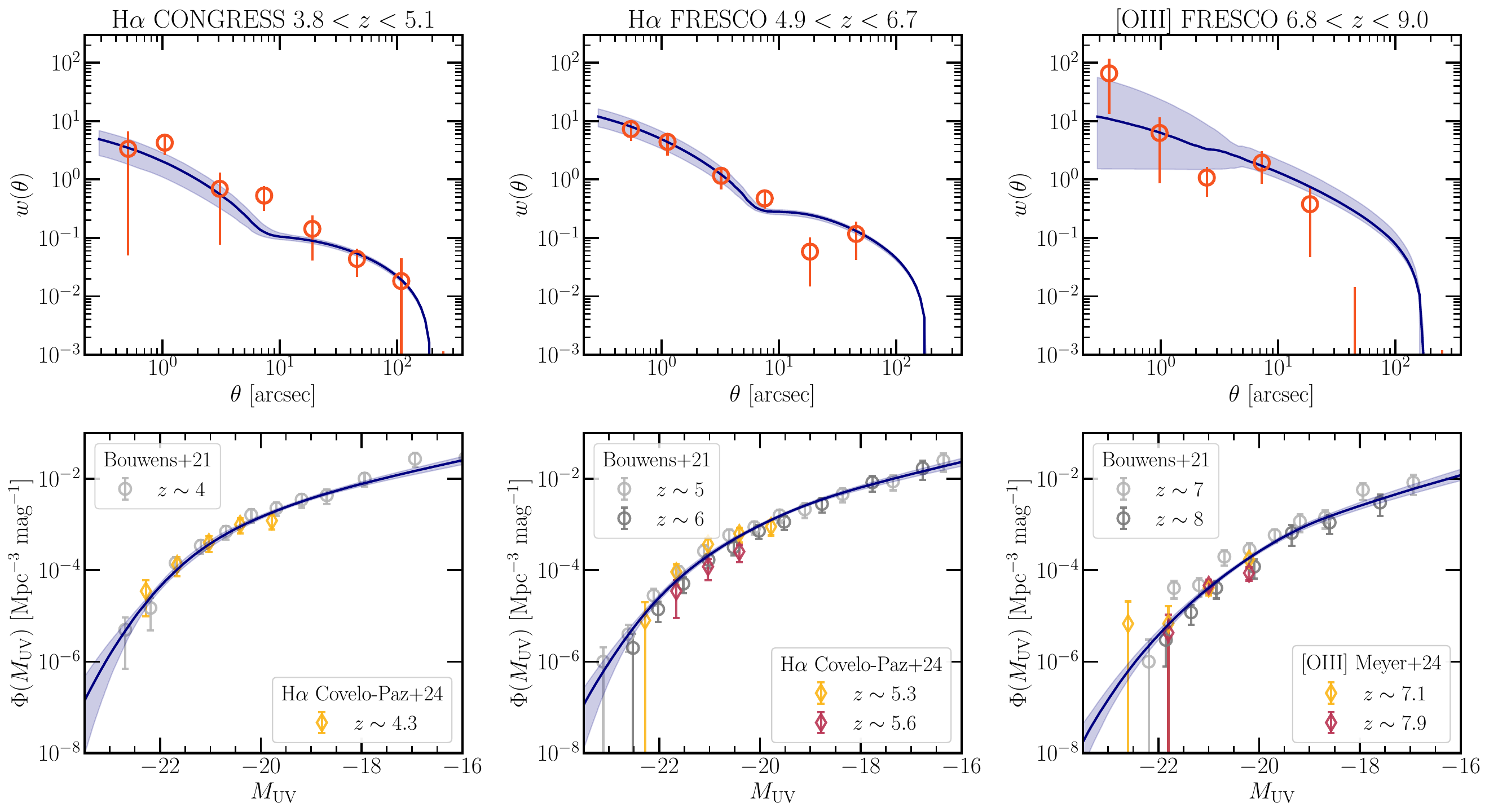}  
\caption{Best fit models of the 2PCF and UVLF. Top row: The angular 2PCF measured from the \ha\ CONGRESS and FRESCO samples at $z\sim4.3$ and $z\sim5.4$ and \oiii\ sample at $z\sim7.3$ (orange points). Bottom row: the UVLF for the same line emitter samples from \cite{Covelo-Paz2024, Meyer2024}, along with the photometrically selected samples from \cite{Bouwens2021}. The blue curves and envelopes show the median models and $1\,\sigma$ uncertainty computed from the posterior. In the case of the UVLF, for simplicity we show only the models in the same redshift bins as the 2PCF measurements. We show the individual fits for each measurement in Fig.~\ref{fig:uvlf-fits-zbins}.}
\label{fig:2pcf-uvlf-fits}
\end{figure*}

We implemented the theoretical framework presented in \S\ref{sec:theoretical-framework} to fit simultaneously the measurements of the 2PCF in the three redshift bins (\S\ref{sec:measurements-2pcf}, \S\ref{sec:clustering-results}) and the ten UVLFs across the $3.8<z<9.0$ range (\S\ref{sec:measurements-uvlf}). In Fig.~\ref{fig:2pcf-uvlf-fits} we show the resulting median (solid lines) and $1\,\sigma$ uncertainty (envelope) computed from sampling the posterior distribution of the model parameters. The top row shows the best-fit models of the 2PCF, while the bottom row shows those of the UVLF at the mean redshift of the corresponding bin (e.g. $\langle z \rangle \sim4.3, \, 5.4$ and $7.3$). We show only one UVLF model for clarity, while in Appendix~\ref{apdx:uvlf-fits}, Fig.~\ref{fig:uvlf-fits-zbins} we show the individual fits for each measurement.

The models show good fits of the data within the uncertainties for the 2PCF and UVLFs for all redshifts. In the case of the 2PCF, at the intermediate scales around the 1-halo to 2-halo transition ($\theta\sim10''$) the data show slightly higher amplitudes than the data, likely due to imperfect modeling of the non-linear and scale-dependent halo bias (cf. \S\ref{sec:2pcf-model}). The overall good fit showcases that our framework can successfully model both the UVLF and 2PCF and derive physical insights, which we present in the following.

\subsection{Mean halo mass and galaxy bias} \label{sec:halomass-bias}

\begin{figure}[t!]
\includegraphics[width=1\columnwidth]{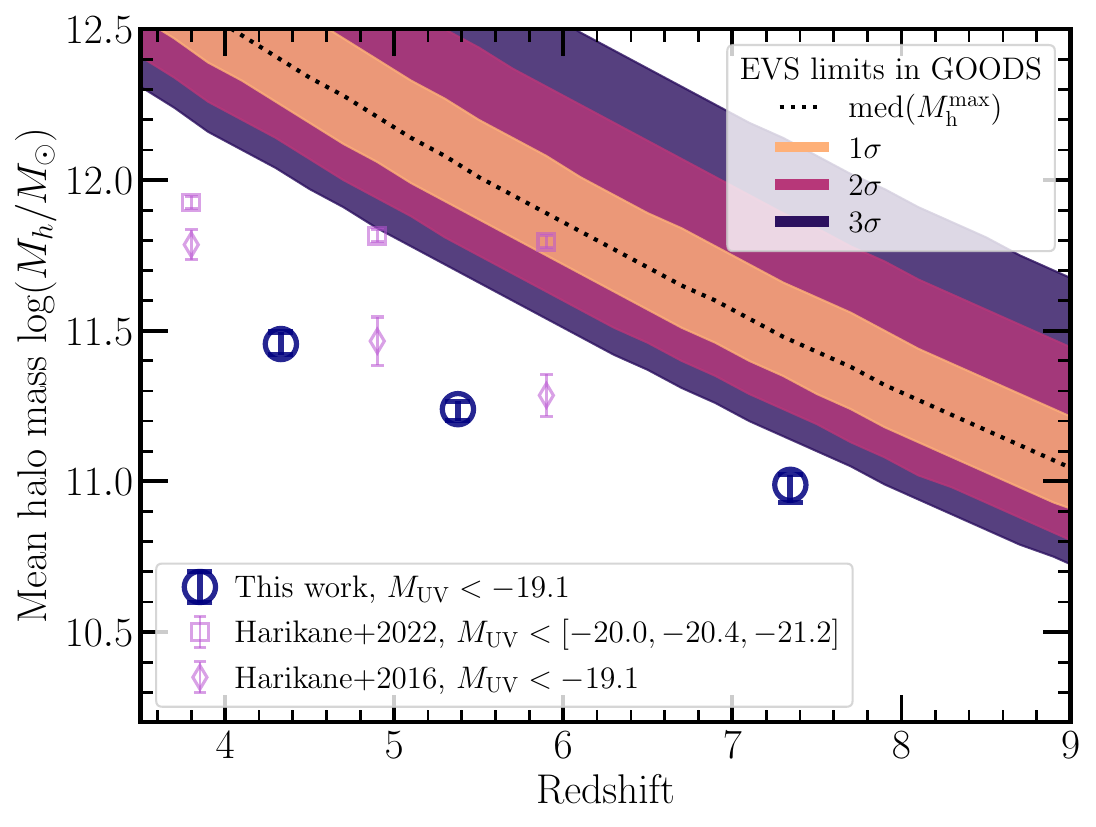}  
\caption{Mean halo mass for the three line emitter samples, \ha\ samples at $z\sim4.3$ and $5.4$, and \oiii\ at $z\sim7.3$ with $M_{\rm UV}<-19.1$ mag. We compare with measurements from \cite{Harikane2016} and \cite{Harikane2022} for photometrically selected LBG. The colored regions are derived from the extreme value statistics \citep[EVS, ][]{Lovell2023} formalism and indicate the confidence intervals to observe the most massive halo in the GOODS volume ($2\times 60\, {\rm arcmin}^2$) within the $\Lambda$CDM model. The dotted line marks the median of the EVS distribution of the maximum plausible halo mass.}
\label{fig:Mhalo-vs-z}
\end{figure}

\begin{figure}[t!]
\includegraphics[width=1\columnwidth]{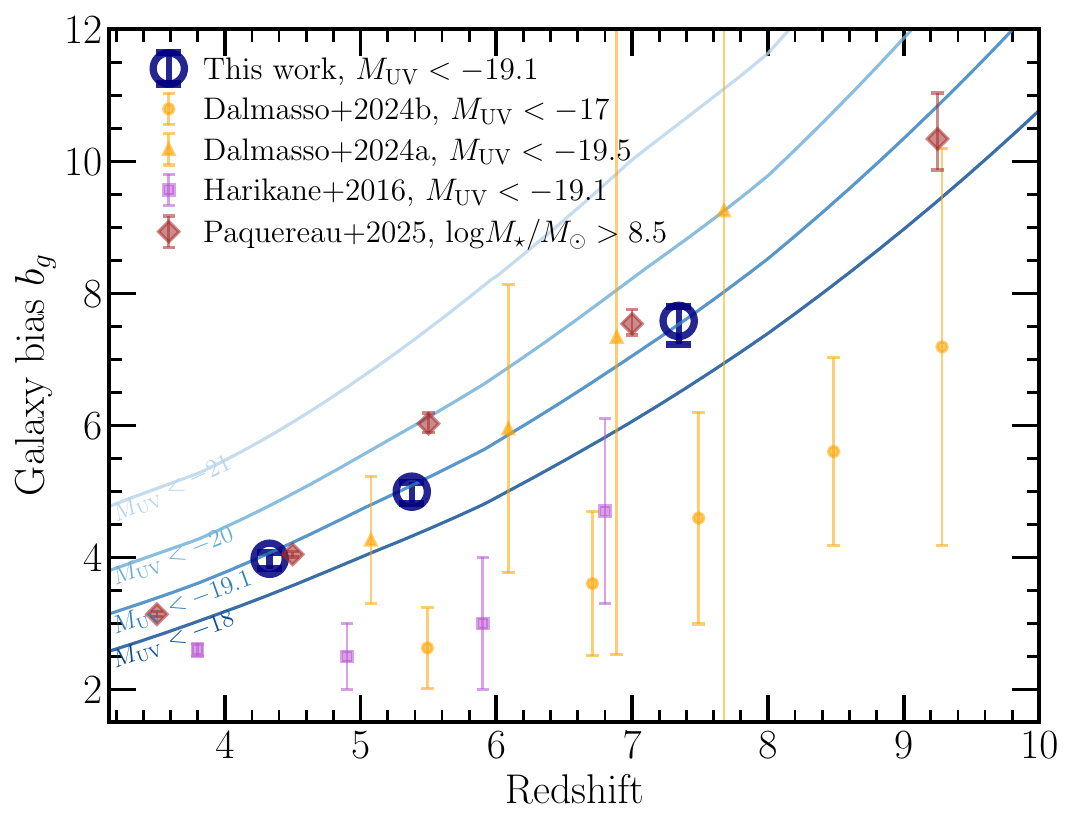}  
\caption{Galaxy bias as a function of redshift and \MUV. The empty circles with error bars show the bias for the three line emitter samples \ha\ at $z\sim4.3$ and $5.4$, and \oiii\ at $z\sim7.3$ with $M_{\rm UV}<-19.1$ mag. The solid blue lines show the bias as a function of redshift for four \MUV\ thresholds computed from our model. We compare with measurements from \cite{Harikane2016}, \cite{Dalmasso2024a} and \cite{Dalmasso24} for photometrically selected LBG, as well as \cite{Paquereau2025} for stellar mass selected, normal galaxies.}
\label{fig:bias-vs-z}
\end{figure}

The HOD modelling of the 2PCF and UVLF allows us to infer the mean mass of the DM halo that host our line emitters samples and their bias (\S\ref{sec:derived-params}). Fig.~\ref{fig:Mhalo-vs-z} and Fig.~\ref{fig:bias-vs-z} shows the mean halo mass and galaxy bias for the \ha\ samples at $z\sim4.3$ and $5.4$, and \oiii\ at $z\sim7.3$.

This analysis shows that the line emitters samples from our work are hosted in halos of log$(M_{\rm h}/$M$_{\odot}) = 11.46_{-0.04}^{+0.04}$, $11.24_{-0.04}^{+0.03}$, $10.99_{-0.06}^{+0.03}$ for the \ha\ and \oiii\ samples at $z\sim4.3, 5.4$ and $7.3$ respectively. These halos are less massive by $\sim0.2$ dex compared to those hosting similarly bright LBG from the \cite{Harikane2016} analysis.

We interpreted the derived halo masses with the extreme value statistics \citep[EVS,][]{Lovell2023} formalism. EVS is a probabilistic approach to estimating the PDF$(M_{\rm h}^{\rm max})$ of observing the most massive halo at a given redshift and given cosmological volume. In Fig.~\ref{fig:Mhalo-vs-z} we show the confidence intervals of the PDF$(M_{\rm h}^{\rm max})$ derived from EVS for our GOODS area ($2\times 60\, {\rm arcmin}^2$). The masses of the halos hosting our line emitter samples are lower than the most massive halo expected in GOODS by $\sim 0.7$ dex. This is unsurprising, given the fact that we measure the mean halo mass hosting a sample of normal, star-forming galaxies with $M_{\rm UV}<-19.1$ mag. We note that the Harikane et al. measurements are made in a much larger volume of the HSC-SSP and are not comparable with these EVS intervals derived for GOODS.

In Fig.~\ref{fig:bias-vs-z} we show the galaxy bias as a function of redshift and \MUV. The blue empty circles correspond to the bias for the three line emitter samples at the corresponding median redshift. We measure a galaxy bias of $b_{\rm g} = 3.98_{-0.15}^{+0.10}$, $5.00_{-0.18}^{+0.14}$, $7.58_{-0.36}^{+0.22}$, for the \ha\ and \oiii\ samples at $z\sim4.3, 5.4$ and $7.3$ respectively.
Since we fitted the continuous redshift dependence of our model, we can derive the redshift evolution of the galaxy bias as a function of \MUV. This is shown in the blue lines in Fig.~\ref{fig:bias-vs-z}; however, we note that this is an extrapolation beyond the $z$ and \MUV\ regime we probed in our work.
Our results show that the bias increases with both redshift and luminosity, as expected from the theory \citep[e.g.][]{Kaiser1984}. Our measurements are comparable those of similarly bright LBGs from \cite{Dalmasso24}. The stellar mass selected sample of \cite{Paquereau2025} also shows comparable values for the bias, although slightly elevated (by $\lesssim 1$) at $5<z<8$. This can be due to differences in the redshift and \MUV\ selection as well as in the modeling.

\subsection{UV magnitude - halo mass relation} \label{sec:MUV-Mh-relation}
\begin{figure}[t!]
\includegraphics[width=1\columnwidth]{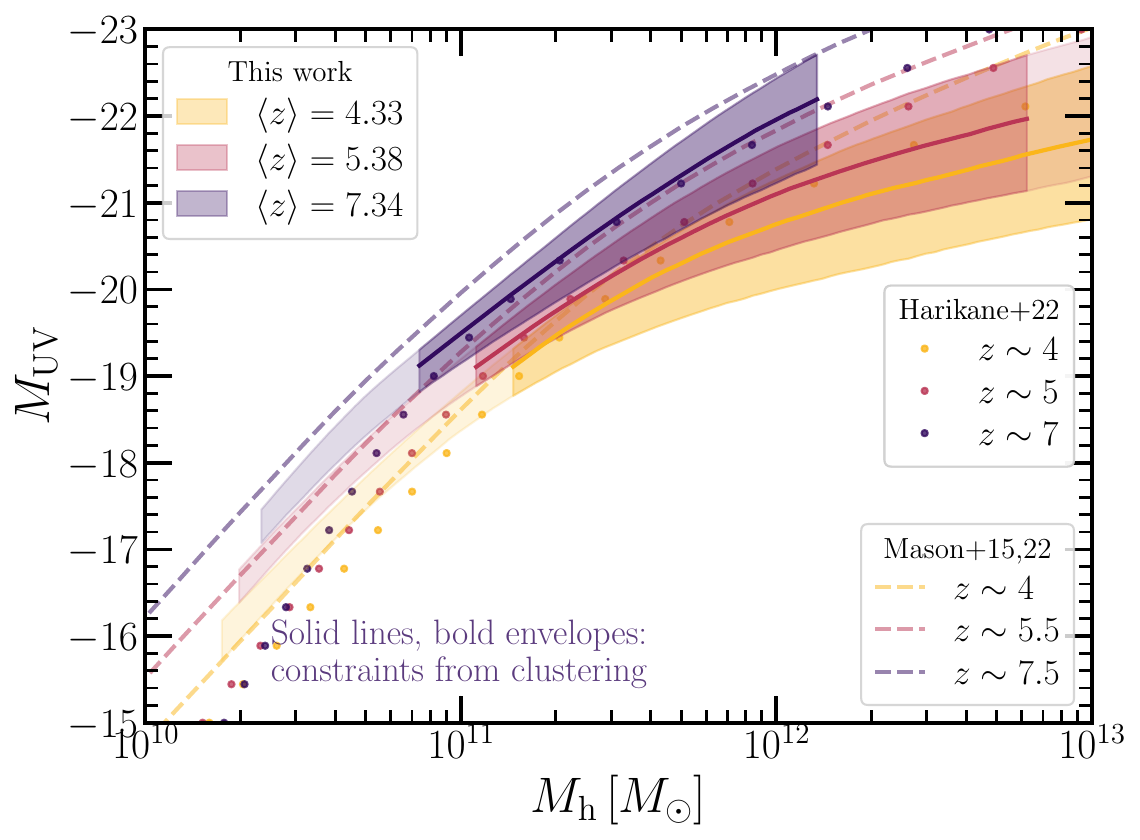}  
\caption{$M_{\rm UV}-M_{\rm h}$ relationship (or UVHMR), for the redshift bins of our three line emitter samples. The solid lines and bold envelopes mark the \MUV\ range that we probe with the 2PCF, while the transparent envelopes mark the range probed by the UVLF. We compare with the model by \cite{Mason2015, Mason2023} that includes dust attenuation in dashed lines and the observational measurements from \cite{Harikane2022} in dots.}
\label{fig:MUV-Mhalo}
\end{figure}

Our framework allows us to measure the \MUV$-$\Mh\ relationship (UVHMR) and its evolution with redshift. In Fig.~\ref{fig:MUV-Mhalo} we show our results on the UVHMR at the median redshifts of our three line emitters samples. However, in principle, since we fit for the redshift evolution of our model parameters, we have information of the continuous evolution of the UVHMR over the redshift range that we probe in this work, as well as extrapolations beyond. In this work, we only probe a limited range in \MUV\ and correspondingly in \Mh\ with the 2PCF, which is marked by the solid lines and bold envelopes in Fig.~\ref{fig:MUV-Mhalo}. The transparent envelopes extend to the minimum \MUV\ that is probed by the UVLF. We show the dust-attenuated \MUV.

The UVHMR shows the typical monotonic increase of UV luminosity with increasing \Mh. As the halo mass increases, the slope of the UVHMR decreases, with a pivot mass at $M_{\rm h}\sim 3\times 10^{11}\, \si{\Msun}$. This is a similar pivot mass scale as the galaxy SHMR \citep{wechsler_connection_2018}. The UVHMR evolves with redshift such that halos of fixed mass host brighter galaxies at earlier epochs. The high mass and luminosity end becomes shallower at lower redshifts due to the more important effect of dust attenuation.
The relatively large uncertainties prevent us from inferring any potential mass or luminosity dependence of the redshift evolution.

We compare our measurements with the theoretical model from \cite{Mason2015, Mason2023} and the observational measurements from \cite{Harikane2022}, which also include dust-attenuation. In the \MUV\ and \Mh\ regime probed by our work, our UVHMR shows shallower slopes for all redshift bins, indicating luminosities that increase faster with halo mass. At fainter luminosities, the constraints from UVLF-only are in better agreement with the \cite{Mason2015, Mason2023} model.

\subsection{The instantaneous star formation efficiency} \label{sec:instSFE}

\begin{figure}[t!]
\includegraphics[width=1\columnwidth]{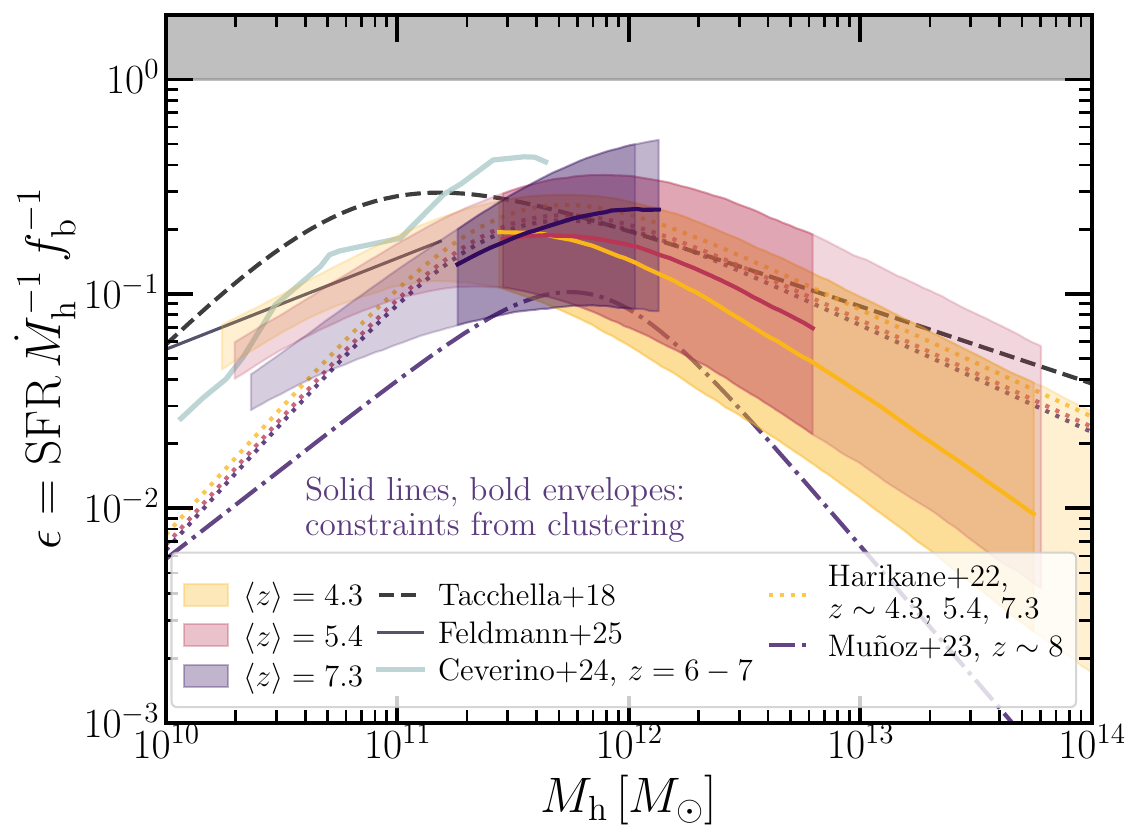}  
\caption{Star formation efficiency as a function of halo mass for the redshift bins of our three line emitter samples. The solid lines and bold envelopes mark the \MUV\ range that we probe with the 2PCF, while the transparent envelopes mark the range probed by the UVLF. We compare with the observational measurements using 2PCF and HOD from \cite{Harikane2022} in dotted lines, empirical model by \cite{Tacchella2018} in  dashed line, and results from cosmological simulations by \cite{Ceverino2024} and \cite{Feldmann2025} in solid colored lines.}
\label{fig:SFE}
\end{figure}

The central feature of our modeling is the parametric form of the instantaneous SFE, and its dependence on halo mass and redshift, $\epsilon(z, M_{\rm h})$ (Eq.~\ref{eq:sfe-parametrization}), that we constrain from observations of the UVLF and 2PCF. As a reminder, $\epsilon$ is defined as the ratio between the SFR and the halo accretion rate times the universal baryon fraction $\epsilon = \dot{M}_{\star} \, \dot{M}_{\rm h}^{-1} \, f_{\rm b}^{-1}$ (Eq.~\ref{eq:sfr-mhdot}), and describes the efficiency of converting gas to stars. 

Figure~\ref{fig:SFE} shows our results on the SFE as a function of halo mass for the three redshift bins of our three line emitter samples. The solid lines and bold envelopes indicate the \MUV\ (and correspondingly \Mh) range probed with the 2PCF, while the transparent envelopes mark the range probed by the UVLF. We note that the upper \Mh\ range is derived from the \MUV-\Mh\ relation, and it can extend to halo masses beyond the maximum mass expected in the GOODS volume, as indicated in Fig.~\ref{fig:Mhalo-vs-z}. Therefore, we caution the interpretation of these results at the highest masses that can be considered as extrapolation.

The SFE shows the characteristic dependence with halo mass -- increasing up to a peak halo mass scale of $M_{\rm h}\sim3\times 10^{11}\, \si{\Msun}$ and decreasing for halos more massive than this peak mass. This characteristic shape of the SFE-\Mh\ relation is typically explained by the effect of stellar and AGN feedback suppressing star-formation at the low- and high-mass end \citep[e.g.][]{silk_current_2012}.

\begin{figure}[ht!]
\centering
\includegraphics[width=0.75\columnwidth]{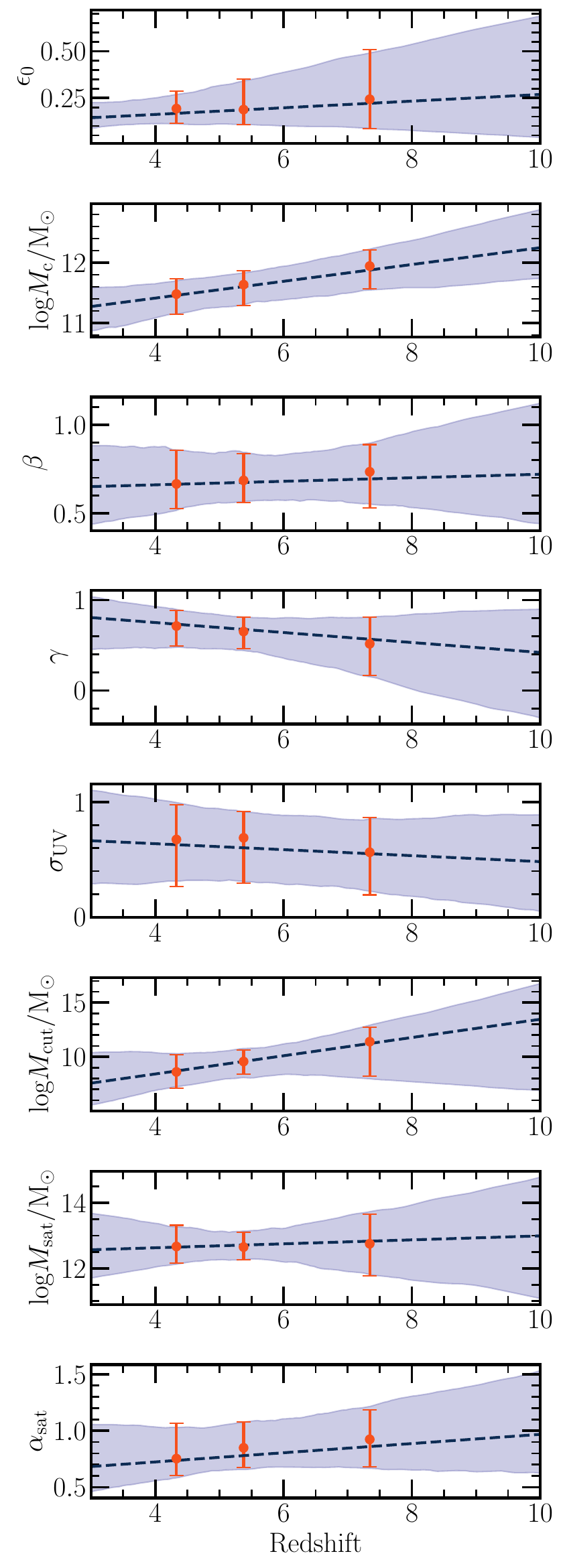}  
\caption{Resulting values of the model parameters. These are obtained by drawing 1000 samples from the posterior and evaluating the models at a given $z$ using Eq.~\ref{eq:param-z-depend}. The orange points and errorbars mark the redshift bins of the 2PCF measurements. The dashed lines mark the $z$-parametrized functions (Eq.~\ref{eq:param-z-depend}) for our model parameters, evaluated at the median posterior.}
\label{fig:best-fit-params}
\end{figure}

As a function of redshift, our results show very little evolution in the $4\lesssim z\lesssim 7$ range. At the low-mass end ($M_{\rm h}<2\times10^{11}\,\si{\Msun}$) the SFE shows a decrease by about 0.3 dex from $z\sim4.3$ to $z\sim7.3$. However, at the high-mass end ($M_{\rm h}>10^{12}\,\si{\Msun}$) there are tentative indications that the trend is reversing, and the SFE increases with redshift, albeit with large uncertainties. This trend becomes more clear when we look at the redshift dependence of the best-fit parameters in Fig.~\ref{fig:best-fit-params}. The normalization of the SFE, $\epsilon_{0}$, shows a mild increase with redshift with a positive slope $\dd \epsilon_0/\dd z = 0.02_{-0.03}^{+0.06}$. However, the lower uncertainty limit is also consistent with zero slope, therefore making it difficult to draw a robust conclusion. The peak mass scale $M_{\rm c}$ also increases with redshift with a slope $\dd {\rm log}M_{\rm c}/\dd z = 0.14^{+0.12}_{-0.10}$, meaning that the SFE shifts towards high masses which creates the decrease of the SFE with redshift at the low-mass end. The high-mass end slope $\gamma$ decreases ($\dd \gamma/\dd z = -0.06_{-0.12}^{+0.10}$), causing the SFE to increase with redshift at a fixed halo mass larger than $M_{\rm c}$. This suggests increasing efficiencies of more massive halos in the early Universe. However, this remains a tentative interpretation, given the fact that the $1\,\sigma$ upper limit on $\dd \gamma/\dd z$ is also consistent with an increasing slope. Additionally, the increase of $M_{\rm c}$ with redshift is primarily driven by the measurements in the $z\sim7.3$ bin, where the 2PCF is relatively noisy. As shown in Fig.~\ref{fig:model-param-dep}, lower clustering amplitude would lower $M_{\rm c}$. These uncertainties highlight the need for larger galaxy samples for more precise clustering measurements in order to conclusively infer the redshift evolution (or lack thereof) of the SFE.

We compare our results with the observational measurements using the 2PCF and HOD modeling from \cite{Harikane2022}, the empirical model by \cite{Tacchella2018} and the results from the cosmological simulations \textsc{FirstLight} \citep{Ceverino2024} and \textsc{FIREbox$^{HR}$} \citep{Feldmann2025}. 
Our analysis is closest to that of \cite{Harikane2022} and the SFE is in relatively good agreement in a 1 dex range around the peak halo mass. At lower ($<10^{11}\,M_{\odot}$) and higher ($>10^{12}\,M_{\odot}$) halo masses,  our measurements indicate higher and lower SFE, correspondingly. 
The \cite{Tacchella2018} empirical model assumes a universal SFE constant with redshift and has a lower peak halo mass by about $0.2-0.3$ dex and higher efficiencies at $<3 \times 10^{11}\,M_{\odot}$.
Compared to the cosmological simulations, \cite{Feldmann2025} find a non-evolving SFE in \textsc{FIREbox$^{HR}$}, although they can only probe the very low mass end due to the limited simulation volume. Their results are in close agreement to ours at $z\sim4.3$, with the difference that our results indicate a mild redshift evolution. Additionally, the slope of their SFE$-$\Mh\ relation is considerably shallower and can have important implications for derived galaxy statistical quantities such as the UVLF and the SFR density, which we discuss in \S\ref{sec:discussion}.
On the other hand, \cite{Ceverino2024} find a SFE that evolves with redshift in \textsc{FirstLight}. Their SFE in the only overlapping $z=6-7$ bin shows elevated efficiencies compared to ours by about $0.2-0.4$ dex, with an indication of a peak occurring at a similar halo mass scale.


\section{Discussion} \label{sec:discussion}

\subsection{Stochasticity versus star formation efficiency} \label{sec:sigma-or-epsilon}

\begin{figure}[t!]
\includegraphics[width=0.99\columnwidth]{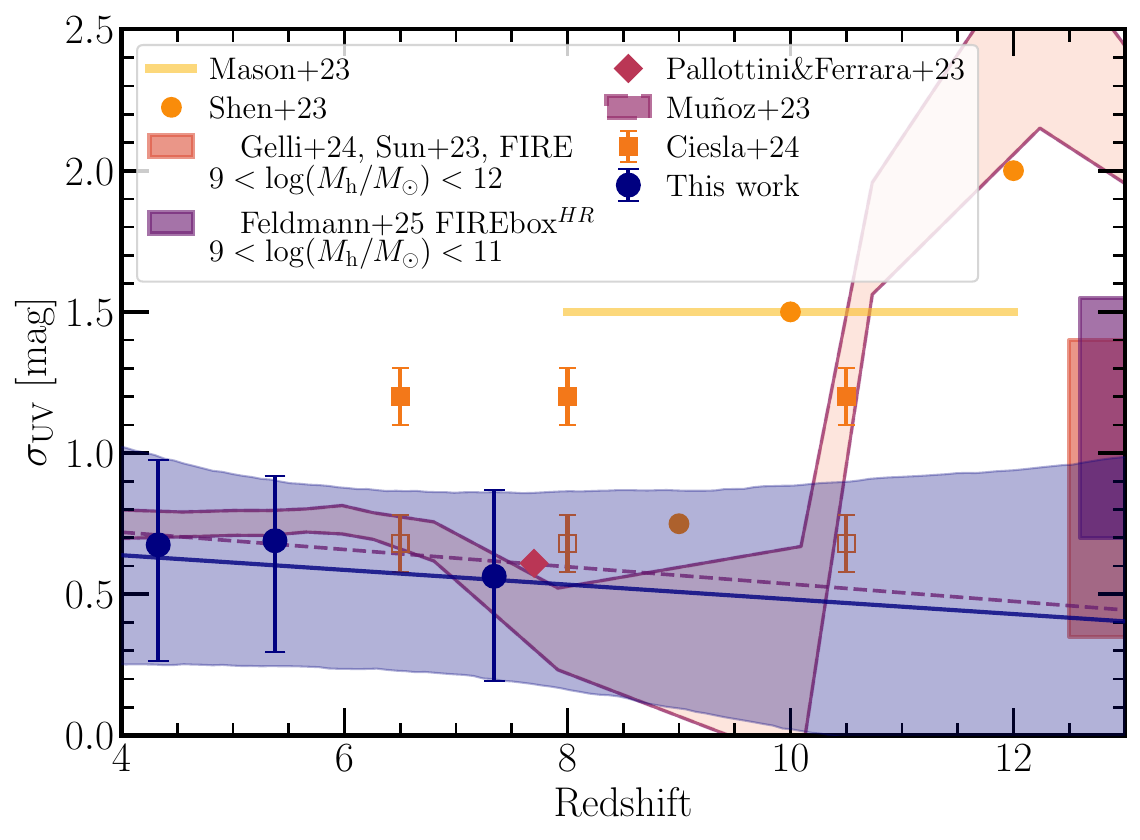}  
\caption{$\sigma_{\rm UV}$, representing the scatter in the \MUV$-$\Mh\ relation, as a function of redshift. The blue points with errorbars mark $\sigma_{\rm UV}$ at the median redshift of our three line emitter samples, while the blue line and envelope mark the best-fit and $1\,\sigma$ uncertainty as a linear function of redshift. From the literature compilation, the colored boxes at the right axis show the $\sigma_{\rm UV}$ for a range of \Mh\ independent of redshift from the \textsc{FIRE} and \textsc{FIREbox$^{ HR}$} simulations \citep{Sun2023, Feldmann2025}. For \cite{Munoz2023}, we show the $1\,\sigma$ contours for $\sigma_{\rm UV}$ obtained for independent $z$-bin fits; while the dashed purple line shows their best-fit linear function with redshift. For \cite{Ciesla2024} we show results from two different methods in the filled and empty squares.
}
\label{fig:sigmaUV-vs-lit}
\end{figure}

\begin{figure*}[t!]
\setlength{\abovecaptionskip}{-0.1mm}
\centering
\includegraphics[width=1\textwidth]{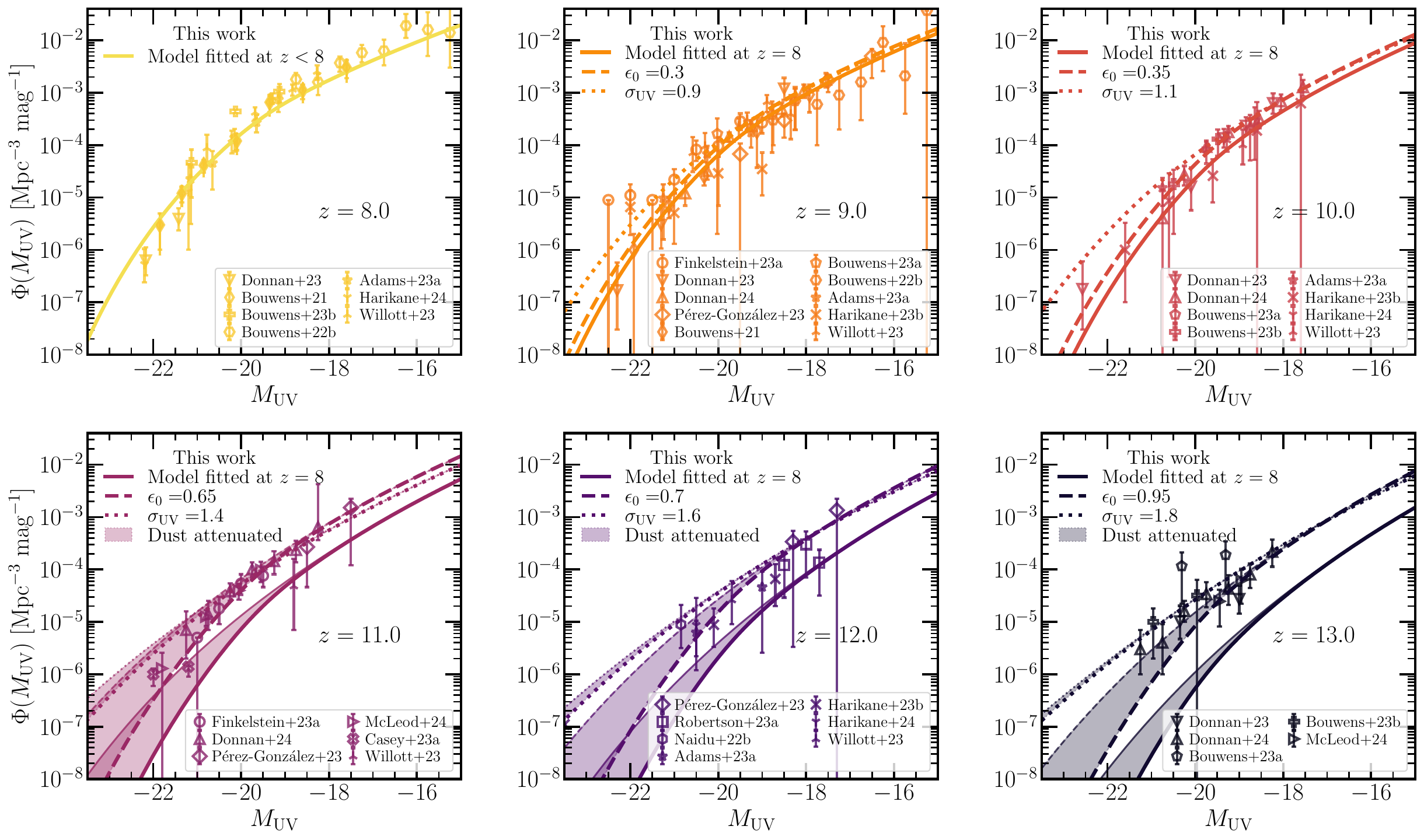}  
\caption{UVLF predictions out to $z=13$ from our model constrained on $z<8$ data and compared to measurements from the literature. The solid lines correspond to the model evaluated at different redshifts from the best-fit linear redshift dependence of the model parameters fitted on $4 \lesssim z \lesssim 8$ data. The dashed and dotted lines illustrate the two extreme scenarios of increasing SFE vs. increasing stochasticity and show the model where we tune by hand $\epsilon_0$ and \sigmaUV\ to approximately match the observations, while keeping the other parameters nominal. The filled area shows the effect of dust attenuation on the UVLF, with the upper thin lines marking the unatenuated UVLF. }
\label{fig:UVLF-high-z-predict}
\end{figure*}


Crucially, our model allows us to break the degeneracy between the high SFE vs. high stochasticity scenarios by quantifying both $\epsilon$ and $\sigma_{\rm UV}$ observationally. In the following, we discuss the implications of our measurements on these two scenarios.

Our results indicate that there is some level of stochasticity, described by modest values of the scatter $\sigma_{\rm UV}\sim0.65$. From our measurements in $4\lesssim z \lesssim 8$ range, we find that \sigmaUV\ remains roughly constant with redshift with a slope $\dd \sigma_{\rm UV}/\dd z = -0.03_{-0.08}^{+0.07}$. 
In Fig.~\ref{fig:sigmaUV-vs-lit} we show the redshift evolution of \sigmaUV\ inferred from our work and compared with the literature. We show in blue points with errorbars our measurements at the mean redshifts of the three line emitter samples. The blue solid line and envelope show the best-fit and $1\,\sigma$ uncertainty as a linear function of redshift. We note that beyond $z\gtrsim8$ these are extrapolations since we do not include any observational (UVLF) constraints.

Compared to the literature, there is a relatively good agreement out to $z\sim9$ from independent approaches. \cite{Munoz2023} fit the UVLF with a similar modeling and find consistent values with ours. Their linear with redshift parametrization of \sigmaUV\ is also consistent with our measurement. However, when they fit for independent $z$-bins they find that \sigmaUV\ needs to increase sharply at $z>10$ to fit the observed \JWST\ UVLFs, a regime currently unconstrained by our framework.

Indeed, there is observational evidence that the transition from secular to stochastic star formation occurs at $z\sim9$ \citep{Ciesla2024, Cole2025, Endsley2024, Looser2023, Dressler2023, Langeroodi2024}. \cite{Ciesla2024}, for instance, study SF histories and show that at $z\sim9$ about $87\%$ of massive galaxies have evidence for a stochastic star formation in the last 100 Myr. However, this is not reflected in the redshift independent $\sigma_{\rm UV} \sim 1.2$ that they infer (filled squares in Fig~\ref{fig:sigmaUV-vs-lit}). This is likely due to the different definition of \sigmaUV\ which is simply the \MUV\ dispersion computed from SED fitting of their sample. Additionally, they estimate \sigmaUV\ in an alternative way by decomposing it into different components and using the SFR$-M_{\star}$ dispersion, and find $\sigma_{\rm UV}\sim0.68$, in agreement with our results (empty squares in Fig~\ref{fig:sigmaUV-vs-lit}).

In the cosmological hydrodynamical simulations \textsc{FIRE-2} \citep{Sun2023} and \textsc{FIREbox$^{HR}$} \citep{Feldmann2025}, \sigmaUV\ is a decreasing function of \Mh\ in the range $\sim 1.40-0.35$ at $9<{\rm log(}M_{\rm h}/\si{\Msun})<12$ for \textsc{FIRE-2}, and $\sim 1.55-0.70$ at $9<{\rm log(}M_{\rm h}/\si{\Msun})<11$ for \textsc{FIREbox$^{HR}$}, both independent of redshift. These are indicated in the colored boxes on the right axis of Fig.~\ref{fig:sigmaUV-vs-lit}. Our results align well with these simulations, as our sample primarily resides in halos with $11 < {\rm log }(M_{\rm h}/\si{\Msun}) <11.5$ (cf. Section~\ref{sec:halomass-bias}), a halo mass regime that would correspond to similar \sigmaUV\ values in the simulations. Finally, in the \textsc{serra} simulation \cite{Pallotini2023} find a \sigmaUV\ at $z\sim7.7$ consistent with ours (diamond symbol in Fig.~\ref{fig:sigmaUV-vs-lit}).

\cite{Gelli2024} discuss the physical motivations behind a halo mass and redshift dependent \sigmaUV. Under the assumption that the stochasticity is driven by the ability of halos to retain the gas despite the feedback processes, then it would scale inversely to the escape velocity, i.e., $\sigma_{\rm UV} \sim v_{\rm esc}^{-1} \sim M_{\rm h}^{-1/3} \, (1+z)^{-1/2}$. This would mean \sigmaUV\ decreases with redshift, which is qualitatively consistent with our measurement of $\dd \sigma_{\rm UV}/\dd z = -0.03_{-0.08}^{+0.07}$.

Our work adds further observational evidence that stochasticity alone cannot explain the observed UVLFs and galaxy clustering measured via the 2PCF out to $z\sim9$. However, the regime in which there is some disagreement with the literature remains at $z>9$, where the transition to stochastic SF is identified. For example, at the highest redshifts ($z>10-16$) \cite{Shen2023} and \cite{Kravtsov2024} find that values as high as $\sigma_{\rm UV}\sim2$ are required to explain the observed UVLF. This is also found by \cite{Munoz2023} as mentioned earlier. This redshift regime is beyond the scope of this work, but future application of our framework to larger datasets extending to $z>9$ promises to provide valuable observational constraints.

We carried out a simple exercise using our theoretical framework to investigate the high stochasticity vs. SFE scenarios at $z>9$. We use the model parameters at $z=8$ which corresponds to the upper redshift limit of our data, and compute the UVLF at $z>8$. Since the parameters fix the HOD at $z=8$, this means that the redshift evolution is primarily driven by the HMF evolution. Figure~\ref{fig:UVLF-high-z-predict} shows the UVLF predictions from our model in solid lines, and compared to measurements from the literature (symbols). Unsurprisingly, at $z\geq9$ our model consistently underpredicts the observed UVLF with the difference increasing with redshift.

We illustrate how two extreme cases would fit the UVLF: 1) increasing only the maximum SFE $\epsilon_0$, and 2) increasing only the stochasticity via \sigmaUV. We tuned both parameters by hand at each $z$ to roughly match the UVLF, while keeping the other parameters to their nominal values (e.g. Fig.~\ref{fig:best-fit-params}). At $z=10$, $\epsilon_0 \sim 35\%$, (compared to the nominal value of $\sim 25\%$) would be required to match the UVLF in case 1, while case 2 requires \sigmaUV$=1.1$, almost double the nominal. At the highest redshifts $z\geq12$ both cases require extreme values of either increasing $\epsilon_0$ to $\sim95\%$, or \sigmaUV\ to $\sim1.8$. 


This exercise shows that neither of the two scenarios alone is likely to explain the observed UVLFs, since they require extreme values for $\epsilon_0$ and \sigmaUV. Such values are typically not predicted by current simulations, although some models \citep[e.g., FFB][]{Torrey2017, Grudic2018, Dekel2023, Li2023, Renzini2023} can produce such high efficiencies.
However, taken together, the impact of different SFEs (e.g., from FFB), and plausible increases in stochasticity, may well result in those high abundances of luminous galaxies.
Therefore, the solution is likely a contribution from both the stochasticity and the SFE along with its dependence on halo mass \citep[e.g.,][]{Gelli2024}. A mass dependence of \sigmaUV\, which we neglect in our work, can have important contributions, but since our framework allows us to implement a halo mass dependence, we will explore this in future work.

To highlight the importance of the SFE dependence on halo mass, we can compare with the results by \cite{Feldmann2025} based on the \textsc{FIREbox$^{HR}$} simulation. They find that a non-evolving, weakly mass dependent SFE (shown in Fig.~\ref{fig:SFE}) can explain the observed UVLF at $z>10$. The reason for this is the shallow slope of their SFE$-$\Mh\ relation at $9< {\rm log}(M_{\rm h}/M_{\odot})<11$ that increases the contribution from the more numerous lower mass halos at higher redshifts -- this can integrate the UVLF to the observed values. This suggests that the global SFE in low-to-intermediate mass halos are higher than previously determined empirically \citep[e.g.][]{Tacchella2018, Harikane2022} or in simulations such as \textsc{FirstLight} \citep{Ceverino2024}. Compared to these models, our empirical results also show shallower SFE$-$\Mh\ slopes at the low mass end, but slightly steeper than \cite{Feldmann2025}, which given the steep increase of the HMF at low masses, could be sufficient to explain why our SFE can not reproduce the observed UVLF at $z>10$.

Higher SFE due to a shallow SFE$-$\Mh\ slope at low-to-intermediate masses is certainly one viable explanation for the observed abundances of bright galaxies at $z>10$. However, it remains to be tested against other galaxy observables such as the 2PCF and SMF and extended at ${\rm log}(M_{\rm h}/M_{\odot})>11$. Indeed, the high-mass regime can be crucial in revealing the full picture. For example, in our case the mild decrease of the SFE with $z$ at the low-mass end is balanced by an increase at the high-mass end and this kind of evolution explains both the 2PCF and UVLF at $4<z<8$. The importance of the bright and massive end in constraining the models and distinguishing between physical mechanisms at $z>10$ is illustrated in Fig.~\ref{fig:model-param-dep} in the context of the SFE model parameters and \sigmaUV. Fig.~\ref{fig:model-param-dep} shows that at $z\sim6$, varying $\epsilon_0 = 0.2 \pm 75\%$ changes $\Phi \, [{\rm Mpc}^{-3}\, {\rm mag}^{-1}]$ by $\sim2$ dex at $M_{\rm UV} = -22$ mag, but $<1$ dex at $M_{\rm UV} = -19$ mag. Similarly, variation of $\sigma_{\rm UV} = 0.6 \pm 75\%$  results in $\Delta \Phi \, [{\rm Mpc}^{-3}\, {\rm mag}^{-1}] \sim 2.5$ dex at $M_{\rm UV} = -22$ mag, but only $\sim 0.2$ dex at $M_{\rm UV} = -19$ mag.

We can conclude from this that it is important to determine the SFE in a large \Mh\ range, spanning several orders of magnitude and beyond the characteristic halo mass of $\sim 10^{12}\, M_{\odot}$ at different epochs in order to understand its (non)evolution with redshift. This can be made possible by implementing our framework to wider and deeper surveys to probe the bright and faint end accordingly, in a wedding-cake approach that can include the whole \JWST\ extragalactic surveys archive. This is necessary in order to provide suffficiently large samples for clustering measurements as well as robust measurements of the bright end of the UVLF.

\subsection{Model degeneracies and how to break them} \label{sec:more-physics}

The framework that we developed in this work is highly predictive, but relies on several physical assumptions that can be degenerate. However, our framework is flexible enough to be extended to model and constrain these against additional observables. 

As mentioned in \S\ref{sec:theoretical-framework}, the luminosity-to-SFR conversion factor $\kappa_{\rm UV}$ is fully degenerate with the maximum SFE, $\epsilon_0$ \citep{Inayoshi2022, Munoz2023, Shen2023}. This means that it is effectivelly the $\epsilon_0/\kappa_{\rm UV}$ ratio that determines the inferred SFE, when constrained against the UVLF. However, $\kappa_{\rm UV}$ might not remain universal at all epochs. For example, different IMF assumption can change $\kappa_{\rm UV}$ from $1.15 \times 10^{-28}\, (\si{\Msun}\, {\rm yr}^{-1})/({\rm erg}\, {\rm s}^{-1})$, for a Salpeter IMF (adopted in this work) to $\kappa_{\rm UV}$ 
to $0.72 \times 10^{-28}\, (\si{\Msun}\, {\rm yr}^{-1})/({\rm erg}\, {\rm s}^{-1})$, for a Chabrier IMF \citep{MadauDickinson2014}. Additionally, lower metallicities and/or top-heavy initial mass functions, as expected for Population III stars, can result in an even lower $\kappa_{\rm UV}$ \citep[e.g.][]{Bromm2002, Inayoshi2022}, that would drive $\epsilon_0$ down in order to reproduce the same UVLF. Finally, $\kappa_{\rm UV}$ depends also on the duration of the previous star formation \citep{Wilkins2019} and on the age of the stellar population, which can also change at earlier times.

This has been implicitly explored in \cite{Donnan2025} who assume an integrated SFE\footnote{Synonymous with the SHMR (\S\ref{sec:uvhmr})} constant with time, tuned to reproduce the SMF at $z\sim 7$, and fit for a $M_{\star}-M_{\rm UV}$ relationship that fits the UVLF at $6<z<13$. The changing mass-to-light ratio evokes younger and thus brighter stellar populations at earlier times which effectively tweaks $\kappa_{\rm UV}$ towards lower values ($\kappa_{\rm UV} \propto 1/L_{\rm UV}$), while forcing $\epsilon_0$ constant. This showcases the $\epsilon_0/\kappa_{\rm UV}$ degeneracy. 

Stellar masses and SMFs provide complementary probes that can help break this degeneracy because they do not depend on $\kappa_{\rm UV}$. This is shown by the fact that the stellar mass is a result of the integrated SFR times a mass loss function\footnote{We note that this is different estimate from e.g., via stellar population synthesis techniques which require an IMF and therefore depend on $\kappa_{\rm UV}$.}, $M_{\star} = \int \dd t \, \dot{M}_{\star} \, (1-f_{\rm r}) $. This also means that the SMF can be modeled under the same parametrization as the UVLF and the 2PCF. $M_{\star}$ being independent of $\kappa_{\rm UV}$ means that the SMF can break the $\epsilon_0/\kappa_{\rm UV}$ degeneracy. However, observational stellar mass estimates depend on the IMF assumption during SED fitting \citep[e.g.,][]{Conroy2009} which can complicate this application.

The SMF is an important probe to consider because accurate measurements of the instantaneous SFE need to be able to reproduce the observed SMF. In the context of the SMF, it is the integrated SFE, $\epsilon_{\star} = M_{\star} \, M_{\rm h}^{-1} \, f_{\rm b}^{-1}$, that describes the relationship between the galaxy stellar mass and host halo mass (SHMR). 
The SHMR has been found to show some level of evolution with redshift by numerous works in the past \citep[e.g.][]{ConroyWechsler2009, Behroozi2010, Moster2010, Moster2013, Behroozi2019_UniverseMachine, girelli2020, Shuntov2022}. Additionally, recent measurements from \JWST\ at $z \gtrsim 7$ using abundance matching and HOD analysis show an important evolution towards high efficiencies at early times \citep{Shuntov2024, Paquereau2025}. 
This seems in contrast with the indications of a very mild evolution of the instantaneous SFE. This suggests that additional modeling components and observational probes (such as the SMF) need to be incorporated in order to accurately inform our models and break degeneracies such as $\epsilon_0/\kappa_{\rm UV}$ and \sigmaUV. Therefore, we identify that the way forward in unveiling a complex picture of intertwined processes is constraining physical models on several complementary probes including the 2PCF, UVLF, SMF and potentially others.

Finally, negligible dust attenuation in the early Universe can also be responsible for the observed abundances of bright galaxies \citep{Ferrara2023, Mason2023}. In our work we used the canonical $A_{\rm UV}-\beta$ \citep{Meurer1999} and $\beta-M_{\rm UV}$ \citep{Bouwens2014} relations to attenuate the \MUV\ when fitting the observed UVLFs, and extrapolating to predict the UVLF at $z>9$. However, this is uncertain because there is evidence from \JWST\ that the $\beta-M_{\rm UV}$ relationship may change towards bluer $\beta$ slopes at $z>8$ \citep[e.g.][]{Cullen2023}, indicating negligible dust attenuation, as well as indications of the opposite \citep[e.g.,][]{Saxena2024}.  We showcase the effect of dust attenuation on the UVLF in Fig.~\ref{fig:UVLF-high-z-predict} with the filled areas, where the upper thin lines mark the unatenuated UVLF. This makes it clear that the level of dust attenuation adds additional degeneracy in the models. Interestingly, in the case of very high stochasticity, the attenuation in the UVLF appears to be negligible. This is unsurprising, because low luminosity galaxies scattered in high mass halos will indeed have little attenuation.
Incorporating the SMF in our formalism would allow for relaxing the assumptions of dust attenuation and fit them directly. However, the difficulty of measuring accurate stellar masses at high redshift is likely to limit that application.

\section{Conclusions} \label{sec:conclussions}

In this paper, we have presented a theoretical framework developed to combine measurements of UVLF and galaxy clustering via the 2PCF into a probe of the galaxy-halo connection. This powerful framework is based on the conditional luminosity function to model the UV luminosity-halo mass relationship (which we called UVHMR) in combination with the HOD. It is a highly flexible adaptation of the HOD framework that allows  the use of independent redshift binning schemes for each probe (UVLF and 2PCF) and parametrize with redshift the model parameters. As such, when fitted on the UVLF and 2PCF it provides empirical constrains on the redshift evolution of the SFE-\Mh\ and \MUV-\Mh\  relationships, stochasticity, host halo masses and galaxy bias. Importantly, this formalism allows for the incorporation of independent probes to test and break degeneracies between different physical models proposed to explain the surprising abundances of bright and early galaxies seen by \JWST.

To implement this framework, we measured the 2PCF of spectroscopically selected \ha\ and \oiii\ line emitters from FRESCO and CONGRESS \JWST\ NIRCam/grism surveys in 124 arcmin$^2$ in the GOODS-North and GOODS-South fields. Our UVHMR$\times$HOD modelling successfully fits the 2PCF and UVLF measurements in three and ten redshift bins at $3.8<z<9$ respectively. Our main findings are summarized below.
   \begin{itemize}
       \item The clustering of $M_{\rm UV}<-19.1$ mag \ha\ and \oiii\ emitters shows 2PCF amplitudes in agreement with similarly bright LBG and normal galaxy samples from literature. The 2PCF amplitude increases mildly from $z\sim4.3$ to $z\sim7.3$ and the shape resembles a power law, requiring modeling the non-linear scale-dependent halo bias, to boost the amplitude at quasi-linear scales ($\sim 100-500$ kpc).

       \item The dark matter halos that host the \ha\ and \oiii\ samples at $z\sim4.3, 5.4$ and $7.3$ have mean masses of log$(M_{\rm h}/$M$_{\odot}) = 11.46_{-0.04}^{+0.04}$, $11.24_{-0.04}^{+0.03}$, $10.99_{-0.06}^{+0.03}$ respectively. The galaxy bias increases with redshift with values of $b_{\rm g} = 3.98_{-0.15}^{+0.10}$, $5.00_{-0.18}^{+0.14}$, $7.58_{-0.36}^{+0.22}$ for the three samples, showing that these galaxies reside in high peaks of the underlying matter density field. However, EVS analysis reveals that their host halos do not correspond to extreme overdense environments for their respective epochs.
    
       \item The SFE$-$\Mh\ relation rises gradually with halo mass, peaks at about $20\%$ at $M_{\rm h} \sim 3 \times 10^{11}\, \si{\Msun} $, and declines at higher halo masses. The SFE$-$\Mh\ shows only a mild evolution with redshift with tentative indications that low mass halos decrease but the high mass halos increase in efficiency with increasing redshift. However, uncertainties are large and consistent with no evolution scenario.

       \item The scatter in the \MUV$-$\Mh\ relationship, quantified by \sigmaUV\ suggests modest stochasticity of $\sim 0.7$ mag, constant with redshift, and in agreement with the literature out to $z\sim9$.

       \item Extrapolating our model at $z>9$ shows that a constant SFE$-M_{\rm h}$ fixed at $z=8$ can not reproduce the observed UVLF. High maximum SFE or high stochasticity alone can not explain the high abundances of luminous galaxies seen by \JWST, since they require extreme and unlikely values (e.g., $\epsilon_0 \sim 0.95$, $\sigma_{\rm UV} \sim 1.8$). However, it is not implausible that the combination of high SFE from FFB and high stochasticity could conspire to result in galaxy formation events that differ from that seen at later times, resulting in an unexpected abundance of highly luminous galaxies at early times.
    
   \end{itemize}

\paragraph{\bf Future prospects.} Extending this analysis to $z>9$ and including additional and complementary probes will be crucial in constraining the galaxy-halo connection and revealing the dominant physical mechanisms in early star formation. Judging from the consistency of our spec-$z$ selected 2PCF measurements with the photo-$z$ selected measurements from the literature, it is promising that photo-$z$ selected samples throughout all \JWST\ surveys will provide sufficiently large samples for 2PCF and UVLF measurements out to the highest redshifts \citep[consistent with the predictions from mocks by][]{Endsley2020}. This will be possible because photometric selection allows the selection of fainter samples, which are more numerous.
However, survey width will be crucial too in constraining the UVLF and clustering of bright galaxies, where the model parameters are more sensitive and can thus shed light on the underlying physical mechanisms.

\section*{Data availability}
To make our results transparent and facilitate comparison we provide all our measurements which can reproduce the figures in this paper in tabulated form at \url{https://github.com/mShuntov/Clustering_and_HOD_of_Ha-and-OIII_emitters}.

\begin{acknowledgements}
      This work is based on observations made with the NASA/ESA/CSA James Webb Space Telescope. The raw data were obtained from the Mikulski Archive for Space Telescopes at the Space Telescope Science Institute, which is operated by the Association of Universities for Research in Astronomy, Inc., under NASA contract NAS 5-03127 for \textit{JWST}. 
      Some of the data products presented herein were retrieved from the Dawn JWST Archive (DJA). DJA is an initiative of the Cosmic Dawn Center, which is funded by the Danish National Research Foundation under grant No. 140 (DNRF140).
      This work was made possible by utilizing the CANDIDE cluster at the Institut d’Astrophysique de Paris, which was funded through grants from the PNCG, CNES, DIM-ACAV, and the Cosmic Dawn Center and maintained by S. Rouberol.
      We thank Giorgos Nikopoulos for kindly sharing a neat UVLF literature compilation. We thank Anne Hutter for the valuable discussions.
      This work has received funding from the Swiss State Secretariat for Education, Research and Innovation (SERI) under contract number MB22.00072, as well as from the Swiss National Science Foundation (SNSF) through project grant 200020\_207349.
      This work was further supported by funding from \textit{JWST}-GO-01895, provided through a grant from the STScI under NASA contract NAS5-03127.
      \\
This work made use of the following Python packages: \texttt{numpy} \citep{harris2020array}, \texttt{matplotlib} \citep{Hunter2007}, \texttt{scipy} \citep{2020SciPy-NMeth}, \texttt{astropy} \citep{Astropy2022}, \textsc{halomod} \citep{Murray2021}

\end{acknowledgements}

%
%

\bibliographystyle{aa}
\bibliography{biblio.bib}

\begin{appendix}
\onecolumn
\section{UVLF, 2PCF and bias dependence on model parameters} \label{apdx:model-param-depend}
\vspace{-1mm}

\begin{figure*}[th!]
\begin{center}
\includegraphics[width=0.8\textwidth, trim=0cm 0.0cm 0cm 0.2cm, clip]{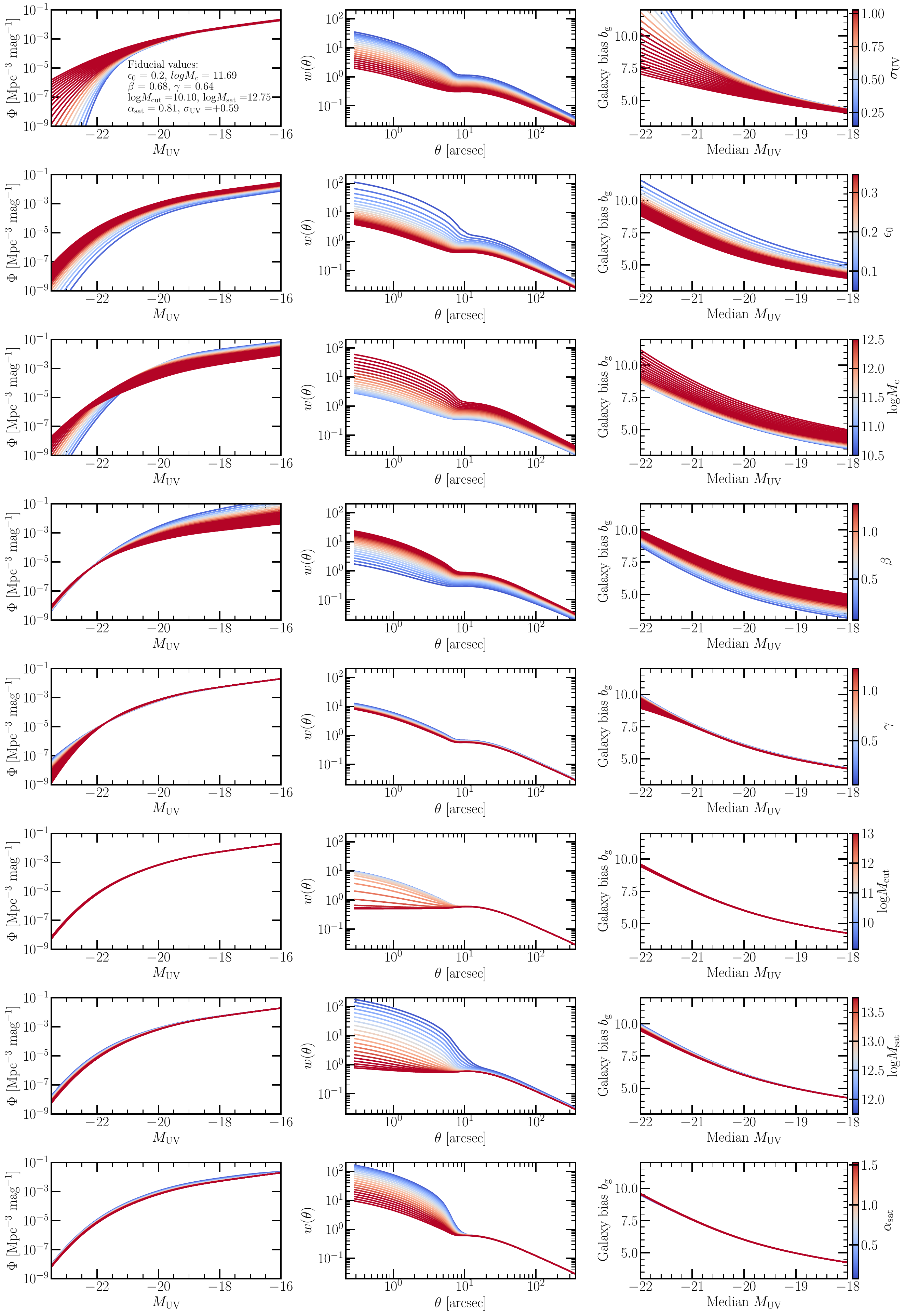}  
\setlength{\abovecaptionskip}{-0.1mm}
\caption{Dependence of the UVLF, 2PCF and galaxy bias at $z=6$ on model parameters. Different rows show the UVLF (left), 2PCF (middle) an bias (right column) for different values of the seven model parameters $\sigma_{\rm UV}, \, \epsilon_{0}, \, {\rm log} M_{\rm c}, \, \beta, \, \gamma, \, {\rm log} M_{\rm cut}, \, {\rm log} M_{\rm sat}, \alpha_{\rm sat}$. One parameter is varied at a time while keeping the others at their fiducial values (shown in the top right panel).}
\label{fig:model-param-dep}
\end{center}
\end{figure*}

In Fig.~\ref{fig:model-param-dep} we show how the UVLF, 2PCF and galaxy bias at $z=6$ change with changing values of each of the model parameters. We vary one parameter at a time while keeping the others at their fiducial values (shown in the top right panel). 

\section{Best-fit parameter values} \label{apdx:best-fit-params}

\begin{table}[ht]
\centering
\begin{tabular}{lccc}
\hline
Main model parameter at redshift & $z=4.3$ & $z=5.4$ & $z=7.3$ \\
\hline
$\epsilon_0$ & $0.19_{-0.08}^{+0.09}$ & $0.19_{-0.08}^{+0.16}$ & $0.24_{-0.16}^{+0.26}$ \\
log$M_{\rm c}/$M$_{\odot}$ & $11.48_{-0.34}^{+0.26}$ & $11.64_{-0.35}^{+0.23}$ & $11.94_{-0.38}^{+0.27}$ \\
$\beta$ & $0.67_{-0.14}^{+0.19}$ & $0.69_{-0.13}^{+0.15}$ & $0.73_{-0.21}^{+0.15}$ \\
$\gamma$ & $0.71_{-0.22}^{+0.17}$ & $0.65_{-0.19}^{+0.16}$ & $0.52_{-0.35}^{+0.29}$ \\
$\sigma_{\rm UV}$ & $0.68_{-0.41}^{+0.30}$ & $0.69_{-0.40}^{+0.23}$ & $0.56_{-0.37}^{+0.30}$ \\
log$M_{\rm cut}/$M$_{\odot}$ & $8.61_{-1.49}^{+1.58}$ & $9.57_{-1.18}^{+1.06}$ & $11.40_{-3.15}^{+1.31}$ \\
log$M_{\rm sat}/$M$_{\odot}$ & $12.67_{-0.51}^{+0.65}$ & $12.65_{-0.39}^{+0.46}$ & $12.75_{-0.99}^{+0.90}$ \\
$\alpha$ & $0.75_{-0.15}^{+0.31}$ & $0.85_{-0.17}^{+0.23}$ & $0.92_{-0.24}^{+0.26}$ \\
\hline
Derived parameters  & & & \\
\hline
log $M_{\rm h}/$M$_{\odot}$ &$11.46_{-0.04}^{+0.04}$ & $11.24_{-0.04}^{+0.03}$ & $10.99_{-0.06}^{+0.03}$ \\ 
$b_{\rm g}$ &$3.98_{-0.15}^{+0.10}$ & $5.00_{-0.18}^{+0.14}$ & $7.58_{-0.36}^{+0.22}$ \\

\hline
\hline
Linear function of redshift parameter values & & & \\
\hline
$d \epsilon_{0}/dz$ &  $0.02_{-0.03}^{+0.06}  $ &  &\\ 
$C(\epsilon_{0})$ &  $0.09_{-0.18}^{+0.15}  $ &  &\\ 
$d {\rm log} M_{\rm c}/dz$ &  $0.14_{-0.10}^{+0.12} $  &  &\\
$C($log$M_{\rm c}/$M$_{\odot}$) &  $10.86_{-0.68}^{+0.57}   $ &  & \\
$d \beta /dz$ &  $0.01_{-0.07}^{+0.07}  $ &  & \\
$C(\beta)$ &  $0.62_{-0.41}^{+0.46}  $ &  & \\
$d \gamma/dz$ &  $-0.06_{-0.12}^{+0.10}  $ &  & \\
$C(\gamma)$ &  $0.97_{-0.58}^{+0.54}  $ &  & \\
$d \sigma_{\rm UV}/dz$ &  $-0.03_{-0.08}^{+0.07}  $ &  & \\
$C(\sigma_{\rm UV})$ &  $0.74_{-0.58}^{+0.62}  $ &  & \\
$d {\rm log} M_{\rm cut} /dz$ & $0.84_{-1.11}^{+0.74}  $ &  & \\
$C({\rm log}  M_{\rm cut}/$M$_{\odot}$) & $5.06_{-4.29}^{+5.52}  $ &  & \\
$d {\rm log} M_{\rm sat} /dz$ & $0.06_{-0.48}^{+0.33}  $ &  & \\
$C({\rm log} M_{\rm sat}/$M$_{\odot}$) & $12.38_{-1.69}^{+2.56}  $ &  & \\
$d \alpha_{\rm sat}/dz$ & $0.04_{-0.09}^{+0.09}  $ &  & \\
$C(\alpha_{\rm sat})$ & $0.56_{-0.47}^{+0.64}  $ &  & \\

\hline
\end{tabular}
\caption{Posterior median and $1\,\sigma$ uncertainty of the model parameters. The first part of the table shows the main model parameter values for different redshift bins. The second part shows the derived parameters: mean halo mass and galaxy bias for the three redshift bins. The third part shows the values for the linear parametrization of redshift for the main model parameters (Eq.~\ref{eq:param-z-depend}).}
\label{tab:parameters}
\end{table}
\section{Fits of the UVLF} \label{apdx:uvlf-fits}

\begin{figure}[th!]
\centering
\includegraphics[width=1\textwidth]{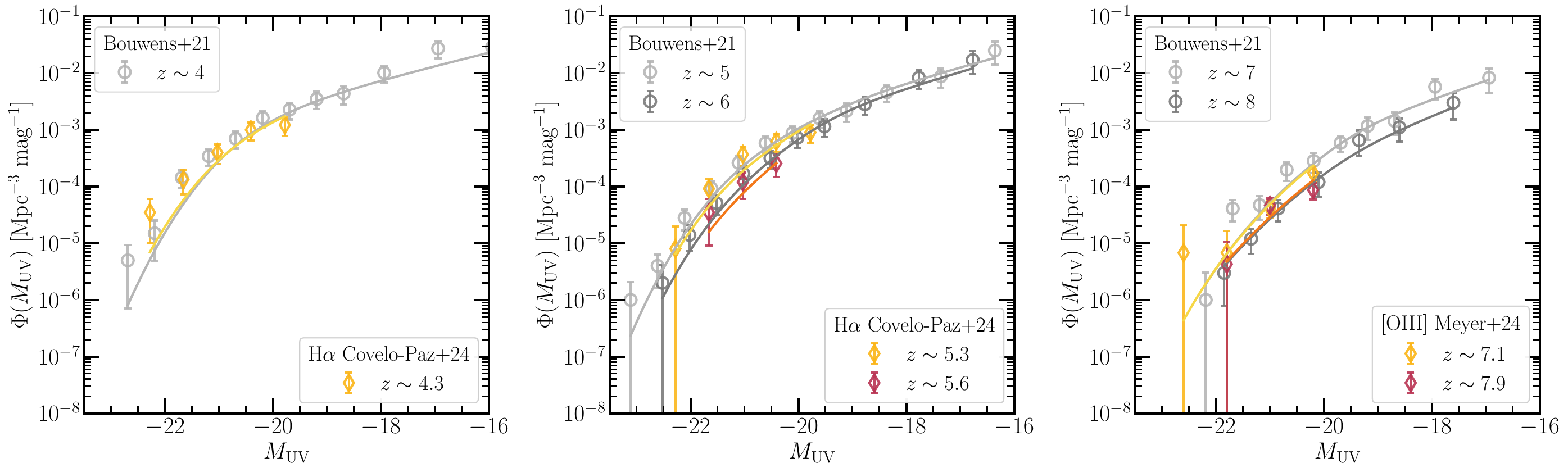}  
\caption{UVLF fits for each individual measurement, color coded accordingly. We plot the models only in the $M_{\rm UV}$ range of the UVLF measurement. These show a good fit between the model and measurements in all $z$-bins.}
\label{fig:uvlf-fits-zbins}
\end{figure}

In Fig.~\ref{fig:uvlf-fits-zbins} we show the UVLF fits for each individual measurement.

\section{Details on the HOD-model derivation of the correlation functions} \label{apdx:woftheta-model}

\subsection{Clustering correlation function}
Briefly described, a central component of the model is the galaxy-galaxy power spectrum that can be separated in contributions from clustering of galaxies within the same halo (1-halo term) and between different halos (2-halo term)
\begin{equation}
    P_{\rm gg}(k,z) = P_{\rm gg}^{1\rm h}(k,z) + P_{\rm gg}^{2\rm h}(k,z).
\end{equation}
These two components of the power spectrum of galaxies can be modelled under the HOD framework \citep{CooraySheth2002, berlind_halo_2002} where the 1-halo and 2-halo terms are given by
\begin{equation} \label{eq:power-spec-1h-2h}
    \begin{split}
        & P_{\rm gg}^{1\rm h}(k,z) = \dfrac{1}{\bar{n}_g^2}\displaystyle\int \dd M_h \dfrac{\dd n}{\dd M_h} \left[ \langle N_s \rangle^2 u_s^2(k) + 2 \langle N_s \rangle \, u_s(k) \right], \\
        & P_{\rm gg}^{2\rm h}(k,z) = \\
        & \dfrac{1}{\bar{n}_g^2} \left[  \displaystyle\int \dd M_h \dfrac{\dd n}{\dd M_h} b_h(M_h,z) \left[ \langle N_c \rangle + \langle N_s \rangle \, u_s(k) \right] \right]^2 P_{\rm lin}(k,z).
    \end{split}
\end{equation}
In these equations, $\bar{n}_g = \int \dd M_h \frac{\dd n}{\dd M_h} (\langle N_c \rangle + \langle N_s \rangle)$ is the mean number density of galaxies, $b_h(M_h,z)$ is the large-scale halo bias which we chose the one given by \cite{tinker_large_2010}. $u_s(k)$ is the Fourier transform of the over-density profile of satellite galaxies, for which we assume that it follows the NFW profile \cite{navarro_universal_1997} with a mass-concentration relation as calibrated by \cite{Duffy08}. The important thing to note here is that the power spectrum is defined in terms of the occupation distributions of centrals and satellites $\langle N_c \rangle$ and $\langle N_s \rangle$ specified by Eq. \ref{eq:hod-ncent} \& \ref{eq:hod-nsat} and is where the HOD parametrization enters the model. The linear power spectrum $P_{\rm lin} (k,z)$ enters the 2-halo term and dominates large scales. Details on deriving the spatial galaxy correlation function $\xi(r)$ and the angular correlation function $w(\theta)$ using \cite{limber_analysis_1953} equation are given in \cite{Blake2008}.

\end{appendix}

\end{document}